\documentclass[twoside]{article}
\usepackage{a4wide}
\usepackage{amsmath}
\usepackage{amsfonts}
\usepackage{amssymb}
\usepackage{times} 

\def\date{July 28, 2003}
\newcommand{\ed}{\end{document}}
\renewcommand{\theequation}{\arabic{section}-\arabic{equation}}
\newcounter{mycnt}[section]
\def\themycnt{\addtocounter{mycnt}{1}\thesection.\arabic{mycnt}}
\def\myb#1{\vskip 3pt\noindent{\bf #1 \themycnt.}}
\def\mye{\hfill\rule{1ex}{1ex}\vskip 3pt}

\def\nn{\nonumber \\}
\def\Id{\text{I\!d}}

\def\openZ{\mathbb{Z}}

\def\openQ{\mathbb{Q}}
\def\antip{{\sf S}}
\def\outerM{{\sf M}}
\def\innerm{{\sf m}}
\def\!{\kern -0.15ex}

\def\Tens{\text{\sf Tens}}
\def\Pleth{\text{\sf Pleth}}
\def\End{\text{\sf End}}
\def\ch{\text{ch}}
\def\ip{\star}
\begin{document}
\title{A Hopf laboratory for symmetric functions}
\author{Bertfried Fauser and P. D. Jarvis}
{ {\renewcommand{\thefootnote}{\fnsymbol{footnote}}
\footnotetext{\kern-15.3pt AMS Subject Classification:
05E05; 
16W30; 
20G10; 
11E57
.}
}}
\maketitle
\begin{abstract}
An analysis of symmetric function theory is given from the perspective
of the underlying Hopf and bi-algebraic structures.  These are
presented explicitly in terms of standard symmetric function
operations.  Particular attention is focussed on Laplace pairing,
Sweedler cohomology for 1- and 2-cochains, and twisted products (Rota
cliffordizations) induced by branching operators in the symmetric
function context.  The latter are shown to include the algebras of
symmetric functions of orthogonal and symplectic type.  A commentary
on related issues in the combinatorial approach to quantum field
theory is given.

\end{abstract}
\section{Introduction}

The symmetric group, and its role in representation theory and the
related symmetric polynomials, is central to many descriptions of
physical phenomena, from classical through to statistical and quantum
domains.  The present work is an initial attempt to synthesize aspects
of symmetric function theory from the view point of the structure
theory of underlying Hopf algebras and bialgebras.  That such a deeper
framework is available is well-recognised (for references see below). 
However, our aim is to exploit the Hopf algebra theory as fully as
possible.  Specifically, our interest is in deformed products and
coproducts, and their characterisation by cohomological techniques in
the context of symmetric functions.

There are several motivations for an approach of this nature.  In the
first place, the symmetric functions provide a concrete arena and
convenient laboratory for the structures of interest suggested by Hopf
algebras.  Furthermore, the Hopf versions of symmetric function
interrelationships confer systematic explanatory
insights, and considerable scope for generalisations.  Last but not
least, it is our intention to provide specific material to an audience
of mathematical physicists and other practitioners, for whom it is
valuable to meld concrete constructs with abstract developments, which
we feel are of abiding importance in several areas of application.

Our own primary motivation for the present study is that symmetric
functions may well serve as a simplified model for combinatorial
approaches to quantum field theory (QFT).  The triply iterated
structure seen in QFT is perfectly mirrored in symmetric functions. 
Questions such as time- versus normal-ordering, and renormalization,
can be framed in abstract combinatorial and algebraic terms (for
references see below).  The precise analogues in symmetric functions
should be concrete and calculable, and may serve as a germ to
understand the more complex QFT setting.  This work provides the
groundwork for such a study.

The outline of this paper is as follows.  An analysis of symmetric
function theory is developed, from the perspective of the underlying
Hopf and bi-algebraic structures.  These are presented explicitly in
terms of standard symmetric function notation (section 2).  The
implications of Laplace pairings for symmetric function operations and
expansions are detailed, and exemplified for the case of Kostka
matrices (section 2).  In section 3, Sweedler's cohomology (for
references see below) is discussed, with emphasis on the analysis of
1- and 2- cochains, cocycle conditions, and coboundaries, and these
are shown to control deformed products via Rota cliffordization
(section 4).  Associative cliffordizations (derived from 2-cocycles,
modulo a 2-coboundary) include cases isomorphic to standard
multiplication (but non-isomorphic as augmented algebras), and also
the Newell-Littlewood product for symmetric functions of orthogonal
and symplectic type.  The relation between the latter and the
standard outer symmetric function algebra is established, using
certain classes of branching operators associated with symmetric
function infinite series, whose properties are discussed.  These
include some of the so-called `remarkable identities' known for these
series from applications to representation 
theory.  Finally, an appendix on the
elements of $\lambda$-rings for the explicit case of symmetric
functions is included.  The paper concludes with a discussion of the 
main results, the outlook for the approach, and further elaboration of the links
between the above constructs and the combinatorial approach to QFT.

\section{The Hopf algebra of symmetric functions}

\subsection{Background and notation}

As mentioned above, we wish to develop aspects of the theory of
symmetric functions wherein the underlying Hopf-, or at least
bialgebraic-, structures, are exploited.  It is well known that the
symmetric functions form a Hopf
algebra \cite{thibon:1991a,scharf:thibon:1994a,rota:stein:1994a,rota:stein:1994b}.
 However, much of the literature uses abstract $\lambda$-ring notation
(see appendix) and is hence not easily appreciated by a practitioner. 
The abstract approach presented in
\cite{malvenuto:reutenauer:1995a,poirier:reutenauer:1995a} is designed
for the framework of quasi-symmetric functions and the permutation
group.  Although some of the present Hopf and bi-algebras are factor
algebras, the technical complications of this much more general
setting obscure the explicit approach to symmetric functions which we
espouse.

In particular, our interest focuses on \emph{cliffordization} and
other \emph{deformed} products and coproducts.  Cliffordization was
introduced by Gian-Carlo Rota and Joel Stein \cite{rota:stein:1994a},
but is related to a Drinfeld twist
\cite{brouder:fauser:frabetti:oeckl:2002a} in the sense of Sweedler
\cite{sweedler:1968a}.  From Sweedler's approach, one learns that many
properties of a deformed product can be characterized by cohomological
methods.  We are going to employ this for symmetric functions.  The
paper of Rota and Stein, \textit{loc cit}, has as its main motivation
the introduction of \emph{plethystic algebras} in a very general
setting.  However, in the language of letter-place superalgebras, it
is again difficult to access concrete calculations in terms
appreciated by practitioners.  The approach by Thibon, Scharf and
Thibon \cite{thibon:1991a,scharf:thibon:1994a} does use the Hopf
structure, however without exploiting Hopf algebra theory (in fact
making only perfunctory use of the additional coalgebra structures). 
For reasons of accessibility, and for a reasonably self-contained
presentation, we therefore introduce Hopf- and bi-algebras for symmetric
functions in a self-consistent manner here.  For a QFT-based account
of combinatorial issues entering into generating functional formulations,
see \cite{brouder:fauser:frabetti:oeckl:2002a,brouder:schmitt:2002a}. At
several points in the sequel, relevant analogies will be pointed out (see
also the concluding remarks).

We use mainly the notation and definitions of Macdonald
\cite{macdonald:1979a}.  The $\openZ$-graded space of symmetric
functions is denoted $\Lambda= \oplus_n \Lambda^n$ where the
$\Lambda^n$ are those subspaces having $n$ variables $(x_{1}, x_{2},
\ldots, x_{n})$.  A partition of an integer is given either as a
non-increasing list of its parts with possible trailing zeros $\lambda
= (\lambda_1,\ldots,\lambda_s)$, using round parentheses, or may be
given as $\lambda=[r_1,\ldots,r_n]= 1^{r_1}2^{r_2}\ldots n^{r_n}$
where the $r_i$ count the occurrences of parts $i$ in $\lambda$.  The
transposed or conjugated partition $\lambda^{\prime}$ is obtained by
mirroring the Ferrer's diagram at the main diagonal.  The length of a
partition $\ell(\lambda)$ is the number of its (nonzero) parts, and
the weight $|\lambda|$ or $\omega_{\lambda}$ is given by the sum of
its parts, $\lambda_{1}+ \lambda_{2}+ \ldots + \lambda_{s} =$ 
$|\lambda|$ $= 1r_1+2r_2+\ldots+nr_n$.  A further way to describe
partitions is given by Frobenius notation.  This gives the length of
`arms' $\alpha_i$ and `legs' $\beta_i$ of a diagram of a partition
measured from the main diagonal.  The length -- number of boxes -- of
the main diagonal is the Frobenius rank $r$ of a partition,
$\lambda=(\alpha_1,\ldots,\alpha_r\mid \beta_1,\ldots,\beta_r)$.  Thus
for example $\lambda = (5,4,2,2,2,1) = [1,3,0,1,1,0,\ldots] = (4,2
\mid 5,3)$.

It is convenient to introduce various bases in the ring $\Lambda$. 
The complete symmetric functions will be denoted as
$h_\lambda=h_{\lambda_1} \ldots h_{\lambda_s}$ with generating
function $H(t)=\sum_n h_nt^n = \prod_{i=1}^\infty (1-x_{i}t)^{-1}$. 
The elementary symmetric functions are defined as
$e_\lambda=e_{\lambda_1} \ldots e_{\lambda_s}$ with generating
function $E(t)=\sum_n e_nt^n = \prod_{i=1}^\infty(1+x_{i}t)$.  Note
that the $h_r$ are represented by a diagram with a single row, and the
$e_r$ as a single column.  For the definition of further Schur
function series, see below.  We need furthermore the monomial
symmetric functions $m_\alpha = \sum_{w\in S_n}
x_{w(1)}^{\alpha_1}\ldots x_{w(r)}^{\alpha_r}$, and the power sum
symmetric functions $p_\lambda=p_{\lambda_1}\ldots p_{\lambda_r}$,
where $p_n=\sum x_{i}^n$.  Note that the $p_\lambda$ form a
$\openQ$-basis only, but see \cite{grosshans:rota:stein:1987a}.  The
most important basis for applications is that of Schur or
$S$-functions, denoted $s_\lambda$.  Schur functions (also homogenous
symmetric polynomials) can be defined via the Jacobi-Trudi
determinantal formulae from the complete or elementary symmetric
functions, or as a ratio of a determinant of monomials with the van 
der Monde determinant (see \cite{macdonald:1979a}).  Occasionally it is
convenient to adopt the Littlewood notation ${\{}\lambda {\}}$ for
the Schur function $s_{\lambda}$.

\subsection{Addition, products, and plethysm}

Reference to the `ring' of symmetric functions amounts to saying that
there are two mutually compatible binary operations.  The
\textit{addition} is the conventional addition of polynomial
functions, and the \textit{multiplication}, the conventional product of
polynomial functions, is the so-called outer product of symmetric
functions.  In terms of Schur functions it is described by the
well-known Littlewood-Richardson rule on the diagrams of the factors:
\begin{align}
\outerM (s_\lambda\otimes s_\mu) \,=\, s_\lambda\cdot s_\mu &= 
\sum_{\nu} C^{\nu}_{\lambda\,\mu}\,s_{\nu}, \qquad s_\lambda\cdot 
s_\mu = s_\mu\cdot s_\lambda,
\end{align}
wherein the dot will sometimes be omitted; the capital $\outerM$ is 
retained to denote the outer product map.  The \emph{unit} for this product
$1_{\outerM}$ is the constant Schur function, corresponding to the
empty or null partition $s_{0}=1$, sometimes just denoted $1$ in the
sequel.  The Littlewood-Richardson coefficients $C^{\nu}_{\lambda \, 
\mu} = C^{\nu}_{\mu \,\lambda } $
may be addressed as a multiplication table, with non-negative integer coefficients 
since they count the number of lattice paths from $\nu$
to $\lambda$ under some restrictions imposed by $\mu$ or \textit{vice versa}.  The
coefficient is zero unless
$\vert\nu\vert=\vert\lambda\vert+\vert\mu\vert$.  

There is another product on symmetric functions, the \emph{inner
product}, denoted by lower case $\innerm$ or $\ip$, which is displayed
most conveniently in the power sum basis
\begin{align}
\label{eq:2-2}
\innerm(p_\lambda\otimes p_\mu) \,=\, p_\lambda \ip p_\mu &= 
\delta_{\lambda\,\mu}z_\lambda p_\lambda
\end{align} 
where $z_\lambda = \prod i^{r_i}\,r_i!$ with 
$\lambda=[r_1,\ldots,r_n]$ (and $z_{(n)}\equiv z_{n} = n$). 
The unit for the inner product can be given in the power sum basis as 
\begin{align}
1_\innerm &:= \sum_n \frac{p_n}{z_n}, \qquad\text{since}\nn
1_\innerm \ip p_i &:= \sum_n \frac{p_n}{z_n} \ip p_i 
               \,=\, \sum_n \frac{\delta_{ni}z_i}{z_n} p_i \,=\, p_i .
\end{align}
Alternatively, the unit reads in the Schur function basis
\begin{align}
1_\innerm &= \sum_{n\ge0} s_{(n)} \,=\, \sum_{n\ge0} h_n .
\end{align}
A variation on the outer multiplication is the notion of symmetric 
function (right or left) \textit{skew} defined for Schur functions by
\begin{align}
s_{\lambda/\mu} & = \sum_{\nu} s_{\nu}C^{\lambda}_{\nu \, \mu} , \qquad
s_{\mu \setminus \lambda} = \sum_{\nu} C^{\lambda}_{\mu \, \nu}s_{\nu} , 
\qquad s_{\lambda/\mu} = s_{\mu \setminus \lambda}
\end{align}
for which necessarily $|\lambda|\ge|\mu|$ (there is no corresponding 
\textit{inner} skew because the corresponding structure coefficients 
for the inner multiplication are totally symmetrical).

Besides addition and the two products, one can define composition or
\emph{outer plethysm}, denoted as $\circ$, on symmetric functions. 
This reads in terms of power sums as
\begin{align}
g \circ p_n = g(x_1^n,x_2^n,\ldots) \quad ( \,=   p_n \circ g), \quad  p_n \circ p_m = p_{nm}. 
\end{align}
Finally for the definition of \emph{inner plethysm} in a Hopf algebra
approach see Scharf and Thibon \cite{scharf:thibon:1994a}.  All
connectivities will play a joint role in what follows.

\subsection{Schur scalar product}

It is convenient to introduce a scalar product on symmetric functions. 
The prominent role which the Schur functions play is reflected in that they 
form an orthonormal basis with respect to the scalar product $(~.~ \mid~.~ )$, by 
definition
\begin{align}
(s_\lambda \mid s_\mu) &= \delta_{\lambda\,\mu}.
\end{align}
Furthermore, the $p_\lambda$ form an orthogonal basis only,
\begin{align}
(p_\lambda \mid p_\mu) &= z_\lambda\,\delta_{\lambda\,\mu}.
\end{align}
with $z_\lambda$ as in (\ref{eq:2-2}) above. The scalar product 
allows us to define dual elements in a unique way\footnote{In any finite 
collection $\oplus_{{\sf finite}}\Lambda^n$ which has to be dense in the 
limit $n\rightarrow \infty$.}. Hence any 
symmetric function $f$ can be expanded into a basis, for example in the Schur 
function basis or the power sum basis
\begin{align}
f &= \sum_\lambda (s_\lambda \mid f) s_\lambda 
   = \sum_\lambda \frac{(p_\lambda \mid f)}{z_\lambda}\, p_\lambda
\end{align}

The scalar product can be used to define adjoints. If $F$ is an
operator on the space of symmetric functions, then we define
$(s_\lambda \mid F(s_\mu)\,) = ( G(s_\lambda)\mid s_\mu )$.
In operator theory $G$ would be denoted as $F^*$, but we will have occasion to use
several generic maps where the adjoints have their own names.

\subsection{Variables versus tensor products\label{sec:2.5}}

We have generally omitted the explicit variables in functions such as 
$s_\lambda$. However, we may reconsider this habit as follows. We have until now used 
\emph{only one} species of variables, namely the linearly ordered set 
$\{x_i\}$. We may collect these into a (possibly infinite) set or 
formal variable $X$. All of the above statements having no variable 
may then be re-read as `...with all variables from the set $X$'. 
However, it turns out that all notions make perfect sense if the symmetric 
functions are considered as \emph{operators} on the formal variable $X$, see
\cite{macdonald:1979a} chapter 1 appendix. This is the so-called 
$\lambda$-ring notation (see appendix)\footnote{The name may origin from $\lambda$-calculus,
where one has a `for all' quantifier establishing exactly the meaning given in
this section.}. However, one should note the important change in the
realm of the statements made in this language. 
We are now ready to introduce a second set of variables $Y$ disjoint from 
$X$, and we can consider symmetric functions on formal sums $X+Y$ or formal
products $XY$, namely the sets ${\{}x_{i},y_{j}{\}}$ or ${\{} 
x_{i}y_{j}{\}}$. It is well known that one can give an isomorphism 
$\theta$ between such multi-variable settings, and tensor products on 
$\End \, \Lambda \cong \Lambda\otimes \Lambda$ since $\Lambda\cong 
\Lambda^*$,
\begin{align}
\theta : F(X)G(Y) &\rightarrow F(X)\otimes G(X) \in \Lambda\otimes\Lambda .
\end{align}
In other words, the $X$ and $Y$ keep track of the tensor slot in a tensor 
product. This is the origin of the letter-place idea promoted in 
\cite{grosshans:rota:stein:1987a}. For our purpose, it is enough to use this
identification to and fro for convenience, and to make contact to the 
literature.

As a further step, we consider tensor products of $\Lambda$ forming the
tensor algebra $\Tens[\Lambda] \equiv \Lambda^\otimes = \oplus_n \Lambda^{\otimes^n}$. Due to the
identification made by the Schur scalar product, we find that endormorphisms
of symmetric functions are elements of $\Lambda\otimes\Lambda$, with the second 
factor seen as dual. The endomorphic product is then composition
\begin{align}
\circ &: (\Lambda\otimes\Lambda)\,\otimes\, (\Lambda\otimes\Lambda)
\,\rightarrow\, (\Lambda\otimes\Lambda), \nn
(G\circ &H)(X) \,=\, G(H(X)).
\end{align}
In terms of letter-place algebras, $\Lambda$ is generated by a single 
alphabet\footnote{Possibly of only a single letter!}
$X$, while $\Lambda^\otimes$ is generated by an infinite collection of disjoint
alphabets, hence $\Lambda^\otimes \cong \Tens\,\Tens[x_1+x_2+x_3+ \ldots]$. 
In this way, using associativity, we can extend
the various structures obtained in $\Lambda$ to $\Lambda^\otimes$. More
technically speaking, $\Lambda^\otimes$ provides a symmetric monoidal
category and the product and coproduct maps $\outerM,\innerm,\Delta,\delta$ (see below) 
are morphisms on $\Lambda^\otimes$. 

\subsection{Inner and outer coproducts}

The canonical extension of the Schur scalar product to tensor powers of
$\Lambda$ is 
\begin{align}
(.\mid.) &: \Lambda^\otimes \,\otimes\, \Lambda^\otimes \rightarrow 
\openZ ,\nn
(.\mid.)\mid_{\Lambda^{\otimes^r}\otimes \Lambda^{\otimes^s}} &= \delta_{rs}
\prod_k (.\mid.)_{k} ,
\end{align}
where $(.\mid.)_{k}$ denotes the scalar product in $\Lambda \otimes 
\Lambda$ applied to the $k$-th factors on each side.

We use this scalar product to dualize the outer and inner products,
and so define the outer coproduct $\Delta$ and the inner coproduct
$\delta$ --once more distinguished by case (notation from
\cite{scharf:thibon:1994a}): 
\myb{Definition}
\begin{align}
(\Delta F \mid G\otimes H) &= (F\mid GH) \nn
(\delta F \mid G\otimes H) &= (F\mid G\ip H)
\end{align}
\mye
\noindent Specifically, inserting a power sum basis, we can read off that
\begin{align}
    \label{eq:theta:co:xytrick}
\Delta\,p_i &= p_i\otimes 1 + 1 \otimes p_i \,=\, \theta(p_i(X+Y)), \nn
\delta\,p_i &= p_i\otimes p_i \,=\, \theta(p_i(XY)),
\end{align}
and hence infer that these properties lift to Schur functions 
$s_{\lambda}$ or generic symmetric functions $f,g$.
Since they are dualized from associative products, these coproducts are 
coassociative, and we can define the iterated coproducts
\begin{align}
\Delta^0 &= \Id,\qquad \Delta^1 \,=\, \Delta, \qquad
\Delta^r \,=\, (\Delta\otimes\Id)\circ\Delta^{r-1}
 \,=\, (\Id\otimes\Delta)\circ\Delta^{r-1},
\end{align}
and analogously for the inner coproduct $\delta^r$. The outer coproduct 
$\Delta$ reads on Schur functions in particular
\begin{align}
\Delta(s_\lambda) &=\sum_\alpha  s_{\lambda/\alpha}\otimes s_{\alpha}
\,=\, \sum_{\alpha\,\beta} C^{\lambda}_{\alpha\,\beta}
s_{\beta}\otimes s_{\alpha}
\end{align}
where $\alpha,\beta$ run over all possible partitions (however
only a finite number of terms contribute). Note, that the Littlewood-Richardson
coefficients now make up the \emph{comultiplication table}\footnote{They should be called
\emph{section} coefficients in this context, and the indices should be arranged
as $C^{\alpha\,\beta}_{\lambda}$. We stay, however, with the
standard convention to prevent possible confusion.}.

It is very convenient to hide the complexity of indexing of coproducts
away via Sweedler notation,
\begin{align}
\Delta(a) &= \sum_{(a)} a_{(1)}\otimes a_{(2)} \nn
(\Delta\otimes\Id)\circ\Delta(a) &= 
\sum_{(a)} a_{(1)}\otimes a_{(2)} \otimes a_{(3)},
\end{align}
where the sum is usually also suppressed. If a distinction between
Sweedler indices is needed, we may use the Brouder-Schmitt convention 
\cite{brouder:schmitt:2002a} that 
\begin{align}
\Delta(a) = \sum_{(a)} a_{(1)}\otimes a_{(2)}, \qquad&\qquad
\delta(a) = \sum_{(a)} a_{[1]}\otimes a_{[2]}, 
\end{align}
keeping track of the type of coproduct involved. If partitions are involved, as
for example in Schur functions, we write simply
\begin{align}
\Delta(s_\lambda) &= s_{\lambda(1)} \otimes s_{\lambda(2)}, \qquad 
\delta(s_\lambda) = s_{\lambda{[}1{]}} \otimes s_{\lambda{[}2{]}}.
\end{align}

Having the scalar product and coproducts in hand, a natural status for the symmetric 
function \textit{skew} product can be recognised\footnote{
Technically, the adjoint of outer product as an element of ${\sf 
End}(\Lambda)$ -- the so-called Foulkes derivative (see Macdonald \textit{loc 
cit}).}
, as the natural action of \textit{dual} elements of $\Lambda$ derived from the outer coproduct:
\begin{align}
    \label{eq:dualaction}
    s_{\mu \setminus \lambda} &= ((\mu|.)\otimes \Id)\circ \Delta (\lambda) 
    = (s_{\lambda(1)}\mid \mu) s_{\lambda(2)} \nn
    &= s_{\lambda(1)}(\mu \mid s_{\lambda(2)}) = 
    (\Id \otimes (\mu|.))\circ \Delta (\lambda) = 
    s_{\lambda/\mu}.
\end{align}
As noted already, for the \textit{inner} coproduct this dual action is identical 
to $\ip$ itself.

\myb{Definition} The \emph{counits} $\epsilon^{\Delta},\epsilon^{\delta}$ of the two
coproducts are
\begin{align}
\epsilon^\Delta(p_\lambda) &:= \delta_{\lambda,0} \\
(\epsilon^\Delta\otimes \Id)\circ
\Delta(p_n) &= \epsilon^\Delta(p_n)\otimes 1+\epsilon^\Delta(1)\otimes p_n
           \,=\, p_n \nn
\epsilon^\delta(p_\lambda) &:= 1 \\
(\epsilon^\delta\otimes \Id)\circ
\delta(p_n) &= \epsilon^\delta(p_n)\otimes p_n \,=\, p_n \nonumber
\end{align}
\mye \noindent
In summary, the unit of the outer product is the constant symmetric 
function $s_{0}$, and the corresponding outer counit is the projection 
onto $s_{0}$; the unit of the inner product is given by the series $H(t)$ at
$t=1$, while the inner counit is given by projecting all power 
sums to 1.

\subsection{Hopf algebra and bialgebra structures}

Since we deal with symmetric products and a self dual space with respect to
the Schur scalar product, we can verify that $\Lambda^\otimes$ is a symmetric
tensor category with trivial braiding\footnote{
Hall-Littlewood symmetric functions
and $q$-Kostka-Foulkes polynomials would be associated with the 
introduction of a nontrivial grade
group (see the concluding remarks below).}, \textit{ie} for $V,W\in \Lambda^\otimes$
\begin{align}
{\sf sw}(V\otimes W) &= W\otimes V .
\end{align}
Given the two products, outer $\outerM$ and inner $\innerm$, and the two
coproducts, outer $\Delta$ and inner $\delta$, it is natural to
investigate which pairs have additional bialgebra or Hopf algebra
structure.  Moreover, it is well known from Hopf algebra theory that
one can form convolution products from a pair of a coproduct and a
product, so we have four possible convolutions and the question 
arises as to which of these convolutions admit antipodes.

\noindent
{\bf Case $\mathbf{I}$:} The outer product and outer coproduct $\outerM, \Delta$:  
\myb{Theorem}
The septuple 
$H=(\Lambda,\outerM,1_\outerM,\Delta,\epsilon^\Delta,{\sf sw},\antip)$ 
is a Hopf algebra (denoted the outer Hopf algebra of symmetric 
functions).
\mye \noindent
{\bf Proof:} We know already associativity, coassociativity and 
unit, counit from which the convolutive unit follows, so we need to show
(i) the compatibility axiom for product and coproduct to form a 
bialgebra, 
(ii) the existence of the antipode.
\begin{itemize}
\item[(i)] Consider the image of $\Delta(s_{\lambda})$ under 
$\theta^{-1}$,
\begin{align} 
s_\lambda(x,y) = \sum_\alpha s_\alpha(x) s_{\lambda/\alpha}(y)
\,=\, \theta^{-1}(s_{\lambda(1)} \otimes s_{\lambda(2)}).
\end{align} 
Computing the following product in two different ways gives
\begin{align}
a)\quad s_\lambda(x,y)s_\mu(x,y) 
&= s_{\lambda/\alpha}(x)s_\alpha(y)s_{\mu/\beta}(x)s_\beta(y) \nn
&= s_{(\lambda/\alpha)\cdot (\mu/\beta)}(x)s_{\alpha\cdot\beta}(y) 
\,=\, 
s_{\lambda\cdot\mu/\alpha\cdot\beta}(x)s_{\alpha\cdot\beta}(y)\nn
b)\quad s_\lambda(x,y)s_\mu(x,y) 
&= s_{\lambda\cdot\mu}(x,y) \nn
&= s_{\lambda\cdot\mu/\rho}(x)s_{\rho}(y) \nn
\Leftrightarrow\quad s_{\lambda\cdot\mu(1)}\otimes s_{\lambda\cdot\mu(2)}
&= s_{\lambda(1)}s_{\mu(1)}\otimes s_{\lambda(2)}s_{\mu(2)}
\end{align}
From $a)=b)$ we can conclude that the product is a coalgebra homomorphism,
and the coproduct is an algebra homomorphism, showing the compatibility
axiom.\\ 
\item[(ii)] We have to show that the antipode ${\antip}$ defined as 
\begin{align}
\sum_\alpha \antip(s_{\alpha})\cdot s_{\lambda/\alpha} 
&= 1_\outerM\circ\epsilon^\Delta(s_\lambda) \,=\, \delta_{\lambda\,0}
\end{align}
exists. This can be done by using a recursive argument as in Milnor and
Moore \cite{milnor:moore:1965a}. From this one obtains in lowest orders 
that
\begin{align}
\antip(s_\lambda) &= (-1)^{\vert\lambda\vert}s_{\lambda^\prime}
\end{align}
From generating functions we know that in $\Lambda$,
\begin{align}
H(t)E(-t) &= 1,\nn
\sum_r (-1)^{n-r} h_r e_{n-r} &= \delta_{n0} 
    \,=\, \sum_r (-1)^{n-r}s_{(r)}s_{(1^{n-r})}, 
\end{align}
which could be extended\footnote{
Here and subsequently certain steps are framed in  $\Lambda^\otimes$
rather than $\Lambda$. The $\mbox{}^{\otimes}$ is 
occasionally omitted by abuse of notation.}
to $\Lambda^\otimes$ via the Jacobi-Trudi
formulae; however, we will take another route.  Using $\lambda$-ring
notation we find with Macdonald (\textit{loc  cit}, pp 29-43) that
in $\Lambda^{\otimes}$ the following holds,
\begin{align}
s_{\lambda}(x,y) &= \sum_\mu s_{\lambda/\mu}(x)s_\mu(y), 
\,=\, \sum_{\mu,\nu} C^\lambda_{\mu\nu}s_{\mu}(x)s_{\nu}(y), \nn
s_\lambda(X+Y) &= \sum_\mu s_{\lambda/\mu}(X)s_{\mu}(Y), \nn
s_\lambda(X-X) &= \sum_\mu s_{\lambda/\mu}(X)s_{\mu}(-X), \nn
s_\lambda(0) &= 
\sum_\mu (-1)^{\vert \mu\vert}s_{\lambda/\mu}(x)s_{\mu^\prime}(x)
\end{align}
from which we obtain the desired result\footnote{
$s_{\lambda}(x,-x)_{n}$ for a finite number of variables can be 
regarded as a symmetric function with compound argument $(x_{i}y_{j})$.
Expanding \textit{wrt} ${\{}y_{j}{\}} = {\{}1,-1{\}}$ evaluates the 
superdimension of representations of $GL(1/1)$ occurring in a 
branching \cite{dondi:jarvis:1981} from 
$GL(n/n)$ to $GL(n) \times GL(1/1)$. These are all two-dimensional 
and typical, unless $\lambda=0$, from which the result follows.}
noting that $s_{(0)}(0)=1$ and 
$s_{\lambda}(0)=0$ for $\lambda\not=(0)$.
\end{itemize}
\mye \noindent
Note that the antipode is related up to a sign factor to the $\omega$-involution of Macdonald, 
which yields just the transpose of the partition, $\omega(s_\lambda) = 
s_{\lambda^{\prime}}$. It is this sign factor which turns the antipode into a M\"obius-like
function, inherited from the underlying poset structure of the lattice of
diagrams. 

Finally, note that the coproduct $\Delta$ may look quite different in other 
bases, \textit{eg} in the
power sums we find as noted already
\begin{align}
\Delta(p_n) &= p_n\otimes 1 + 1 \otimes p_n,\qquad \Delta(1)\,=\,1\otimes 1,\nn
\Delta(p_n^2) &= \Delta(p_n)\Delta(p_n) 
\,=\, p_n^2\otimes 1 + 2p_n\otimes p_n + 1\otimes p_n, \nn
(\Delta\otimes \Id)\Delta(p_n)&= p_n\otimes 1\otimes 1 +
1\otimes p_n\otimes 1+ 1\otimes 1\otimes p_n.
\end{align}
Generally
\begin{align}
\Delta(p_n) &= \sum^r_{k=0} \frac{r!}{k!(r-k)!} p_n^k\otimes p_n^{r-k}
\,=\, \sum^r_{k=0} \left( \! \begin{array}{c} r \\ k \end{array} \! \right) p_n^k\otimes p_n^{r-k} \nn
\Delta^{(l-1)}(p_n) &= \sum_{\sum k_i=r} \frac{r!}{k_1!\ldots k_l!}
p_n^{k_1}\otimes p_n^{k_2} \otimes \ldots \otimes p_n^{k_l} 
\end{align}
\noindent
{\bf Aside:} Relations of this type are quite common in `finite operator calculus' \cite{rota:etal:1975a}. 
The above `addition theorem' (see \cite{macdonald:1979a}, p. 43) is an 
analogue of so-called Appell and Scheffer sequences.
\mye \noindent
{\bf Case {$\mathbf{I\!I}$}:} The outer product and inner coproduct $\outerM, \delta$:  
\myb{Theorem}
The algebra $A=(\Lambda,\outerM,1_\outerM)$  and the coalgebra $C=(\Lambda,\delta,\epsilon^\delta)$
form a bialgebra, but not a Hopf algebra.
\mye \noindent
{\bf Proof:}
We compute firstly the homomorphism property for the bialgebra,
\begin{align}
a)&\quad \delta\circ \outerM(p_n\otimes p_m) = \delta(p_{nm}) 
\,=\, p_{nm} \otimes p_{nm} \nn
b)&\quad    \outerM\otimes \outerM(\Id  \otimes {\sf sw}\otimes \Id  )
              (\delta\otimes \delta)(p_n\otimes p_m)\nn
&\qquad\,=\, \outerM\otimes \outerM(p_n\otimes p_m \otimes p_n\otimes p_m) \nn
&\qquad\,=\, p_{nm} \otimes p_{nm}
\end{align}
which shows $a)=b)$ as needed. The antipode has to fulfil 
\begin{align}
{\antip}(p_n)p_n &= 1_\outerM\circ \epsilon^\delta(p_n) \,=\, \sum_{m} \frac{p_m}{z_m} .
\end{align}
Since the righthand side contains terms of all grades $m$, but the
left hand side terms only have grades which are multiples of $n$, and
since the power sums are independent, this requirement cannot be
fulfilled.  \mye \noindent 
{\bf Case $\mathbf{I\!I\!I}$:} The inner product and
outer coproduct $\innerm, \Delta$: \myb{Theorem} The algebra
$A=(\Lambda,\innerm,1_\innerm)$ and the coalgebra
$C=(\Lambda,\Delta,\epsilon^\Delta)$ form a bialgebra, but not a Hopf
algebra.  
\mye \noindent 
{\bf Proof:} We compute firstly the
homomorphism property for the bialgebra,
\begin{align}
a)&\quad \Delta\circ \innerm(p_n\otimes p_m) = \Delta(\delta_{nm}z_n p_n) 
\,=\, \delta_{nm}z_n(p_n\otimes 1+ 1\otimes p_n) \nn
b)&\quad    \innerm\otimes \innerm(1\otimes {\sf sw}\otimes 1)
              (\Delta\otimes \Delta)(p_n\otimes p_m)\nn
&\qquad\,=\, \innerm\otimes \innerm(p_n\otimes p_m \otimes 1\otimes 1 +
  p_n\otimes 1 \otimes 1\otimes p_m + 1\otimes p_m\otimes p_n \otimes 1 
+ 1\otimes 1\otimes p_n\otimes p_m) \nn
&\qquad\,=\, z_n\delta_{nm}p_n\otimes 1 + z_n\delta_{nm}z_n 1\otimes p_n
\end{align}
which shows $a)=b)$ as needed. The antipode ${\antip}^{\Delta,\innerm}$ has to fulfil the following requirement
\begin{align}
{\antip}(p_n)\ip 1+{\antip}(1)\ip p_n &= 1_{\innerm}\circ\epsilon^\Delta(p_n).
\end{align}
Firstly let $n=0$, then 
the righthand-side reduces to $1_{\innerm}$, while the left hand side is 
$2\,\antip(1)1$, which implies that $2\,\antip(1)=1_{\innerm}=\sum_{n}p_n/z_n$. 
Hence we find in the case $n\not=0$
\begin{align}
\antip(p_n)\ip 1+ \frac 12 1_{\innerm}\ip p_n &= \delta_{n0}\sum_m \frac{p_m}{z_m}
\end{align} 
which cannot be fulfilled.
\mye  \noindent
{\bf Case $\mathbf{I\!V}$:} The inner product and inner coproduct $\innerm, \delta$:  
\myb{Theorem}
The coalgebra $C=(\Lambda,\delta,\epsilon^\delta)$ and the algebra 
$A=(\Lambda,\ip,1_\innerm)$ do \emph{not} form a Hopf algebra, and not 
even a bialgebra.
\mye \noindent
{\bf Proof:} To show it is not Hopf, it suffices to see that there does not 
exist an antipode. We use the power sum basis. Assuming the antipode is 
described by the operator
$\antip(p_n) =\sum_m\antip_{nm}p_m$, we can compute
\begin{align}
\antip(p_n) \ip p_n &= 1_\innerm\circ \epsilon^\delta(p_n) , \nn
\sum_m\antip_{nm} p_m \ip p_n &= \sum_m \frac{1}{z_m}\,p_m , \nn
z_n\antip_{nn} p_n &= \sum_m \frac{1}{z_m}\,p_m ,
\end{align}
which cannot be fulfilled due to the linear independence of the 
power sum functions. To show that the structure is not a bialgebra we have to show that this pair of product
and coproduct is not mutually homomorphic, for example
\begin{align}
a)&\quad\delta\circ \innerm(p_n\otimes p_m) = \delta (\frac{\delta_{nm}}{z_n}p_n)
 \,=\, {\delta_{nm}}{z_n}p_n \otimes p_n \nn
b)&\quad    \innerm\otimes \innerm(\Id\otimes {\sf sw}\otimes \Id  )
              (\delta\otimes \delta)(p_n\otimes p_m)\nn
&\qquad\,=\,\innerm\otimes \innerm (p_n\otimes p_m \otimes p_n\otimes p_m)
\,=\,{\delta_{nm}}{z^2_n}(p_n \otimes p_n)
\end{align}
One cannot fulfil $a)=b)$ due to the fact that the coproduct (or the
product) would need to be rescaled with $\sqrt{z_n}$, which drops out of the
ring of integers, or even the quotient field $\openQ$ 
of rational numbers.
\mye
\noindent
{\bf Aside:} The lack of having an antipode in this case may not be so surprising
as at first sight.  Remember that
$\theta^{-1}(\delta(p_\lambda(X)))=p_\lambda(XY)$.  Hence the antipode
would require $Y=1/X$, which is beyond the ring $\Lambda$ of symmetric
functions.  The introduction of $X^{-1}$ as formal variables is
considered for example in Crapo and Schmitt \cite{crapo:schmitt:2000a}
or Borcherds \cite{borcherds:1997a}.  The present problem may hence be
dubbed the `localization problem' of the inner product, in analogy with
the `localization' process of enlarging a ring to its quotient field
in algebraic geometry.  Using divided powers may provide a cure 
(see \cite{grosshans:rota:stein:1987a}).

\begin{table}[h]\begin{center}
\caption{Mutual product-coproduct homomorphisms}
\begin{tabular}{lrl}
\\\hline\hline
$\mathbf{I}$ & $(\Delta,\outerM,{\sf S})$ & outer Hopf algebra  \\
$\mathbf{I\!I}$ & $(\delta,\outerM)$ & bialgebra  \\
$\mathbf{I\!I\!I}$ & $(\Delta,\innerm)$ & bialgebra  \\
$\mathbf{I\!V}$ & $(\delta,\innerm)$ & inner convolution   \\
\hline\\
\end{tabular}\end{center}
\end{table}
The cases $\mathbf{I}$ -- \mbox{$\mathbf{I\!V}$} are summarized in Table 1. 
Evidently the presence of inner (co)products decreases the 
compatibility with Hopf
algebra axioms. One can think about a slightly altered definition of an 
\emph{inner antipode} which would cure this, and allow four 
mutually related Hopf algebras\footnote{The candidates we have in mind are
$\zeta$-functions and M\"obius functions, where the M\"obius function
replaces the antipode. This still fits into the theory of $\lambda$-rings 
\cite{knutson:1973a}.}. 

\subsection{Scalar product -- Laplace pairing}
A pairing is in general a map ${\mathbf \pi} : A\otimes B \rightarrow C$. 
Particular pairings are actions $\bullet : G\otimes M \rightarrow M$, 
multiplications ${\mathbf \mu} : A\otimes A \rightarrow A$ or 
evaluations ${\sf ev} :A\otimes A \rightarrow \openZ$. A pairing which is compatible with a 
coproduct, \textit{ie} forming a bialgebra, is called after Sweedler a 
measuring (for a theory of pairings and measurings see \cite{schmitt:1999a}). We will use the 
scalar-valued case in the present work only, but note some more 
general cases. 
\myb{Definition} A pairing is called a \textit{Laplace} pairing \cite{grosshans:rota:stein:1987a} 
if it enjoys the following two properties:
\begin{align}
i)\quad (w\mid a \cdot b) &= (w_{(1)}\mid a)(w_{(2)}\mid b) \nn
ii)\quad (a\cdot b\mid w) &= (a\mid w_{(1)}) (b\mid w_{(2)}) 
\end{align}
\emph{and} if the product and coproduct are mutually homomorphisms. \mye 
\noindent
The name stems from the fact that these identities imply the expansion rules
for determinants in exterior, and permanents in symmetric, algebras. 
From the definitions, the Schur scalar product generalized to
$\Lambda^\otimes$, enjoys this crucial property with respect to the outer 
Hopf algebra:
\myb{Theorem} The Schur scalar product is a $\openZ$-valued Laplace pairing with respect to
the outer product and coproduct:
\begin{align}
     ( {\{}
     \lambda {\}}\mid  {\{}\mu {\}} \cdot  {\{}\nu {\}}) &= 
                          ( {\{}\lambda_{(1)} {\}}\mid  {\{}\mu {\}})( {\{}\lambda_{(2)} {\}}\mid  {\{}\nu {\}}),
      \quad (  {\{}\mu {\}} \cdot  {\{}\nu {\}}\mid  {\{}\lambda {\}}) = 
                ( {\{}\mu {\}}\mid  {\{}\lambda_{(1)} {\}}) ( {\{}\nu 
                {\}}\mid  {\{}\lambda_{(2)} {\}}). \nonumber
\end{align}
\mye \noindent
{\bf Proof:} This follows from the fact that the outer 
coproduct was introduced by duality from the outer 
product, and that ($\Delta,{\outerM}$) form a bialgebra (cases \mbox{$\mathbf 
{I}$}). 
\mye \noindent 
Note that the corresponding property for inner product and coproduct 
does not constitute a Laplace pairing because of the lack of 
compatibility (Table 1, case \mbox{$\mathbf {I\!V}$}).
From the coalgebra structures, cases \mbox{$\mathbf {I}$},
\mbox{$\mathbf {I\!I}$}, \mbox{$\mathbf {I\!I\!I}$}, more general Laplace 
properties can be inferred, for \textit{non-scalar} pairings, given here for completeness:
\myb{Theorem}
\begin{itemize}
    \item[(i)] The skew product satisfies the (partial) Laplace 
    conditions with respect to the outer and inner products:
    \begin{align}
      ( {\{}\lambda {\}} \cdot {\{}\mu {\}}) / {\{}\nu {\}} &= 
           ({\{}\lambda {\}}/ {\{}\nu_{(1)} {\}})\cdot ({\{}\mu {\}}/ {\{}\nu_{(2)}{\}}),
        \, (  {\{}\lambda {\}} \ip  {\{}\mu {\}})/ {\{}\nu {\}} = 
          ( {\{}\lambda {\}} /  {\{}\nu_{(1)} {\}}) \ip ({\{}\mu 
          {\}}/{\{}\nu_{(2)}{\}}). \nonumber
      \end{align}
      \item[(ii)] The inner product is a $\Lambda^{\otimes}$-valued 
      Laplace pairing with respect to the outer product:
          \begin{align}
       {\{}\lambda {\}} \ip  ({\{}\mu {\}} \cdot {\{}\nu {\}}) &= 
          \sum( {\{}\lambda_{(1)} {\}} \ip  {\{}\mu {\}}) \cdot 
          ({\{}\lambda_{(2)} {\}}\ip{\{}\nu{\}}) , \nn
	  ({\{}\lambda {\}} \cdot {\{}\mu {\}}) \ip {\{}\nu {\}} &= 
           \sum ({\{}\lambda {\}}\ip {\{}\nu_{(1)} {\}})\cdot ({\{}\mu 
           {\}}\ip {\{}\nu_{(2)}{\}}). \nonumber
      \end{align}
\end{itemize}¥
\mye \noindent {\bf Proofs:} Part (i) refers to the outer Hopf 
algebra (case $\mathbf{I}$) and the inner coproduct/outer product 
bialgebra (case $\mathbf {I\!I}$), interpreted for the dual action 
(\ref{eq:dualaction}) -- note however that $s_{\lambda/\mu} \ne 
s_{\mu/\lambda}$ in general. Part (ii) applies directly for the dual 
action from the outer coproduct-inner product bialgebra (case $\mathbf {I\!I\!I}$).
Finally, note that the inner coproduct-inner product convolution 
algebra does not admit the Laplace property.
\mye \noindent
Of course, these formulas are known
\cite{littlewood:1940a,thibon:1992,baker:1994}.  However, the above
reasoning shows that they emerge from a single principle, which in turn
generates Wick-like expansions (see
\cite{fauser:2001b,fauser:2002c,luque:thibon:2002a}, where such
expansions are treated in detail;  for the fermion-boson correspondence see 
\cite{baker:1995a} and \cite{jarvis:yung:1993, 
hamel:jarvis:yung:1997} for diagram strip decompositions and 
determinantal forms).  The amazing computational power of the
Laplace identities cannot be underestimated.

\subsection{Kostka matrices}

An immediate consequence of the observation that the Schur scalar product 
is a Laplace pairing is the fact, that it allows us to give a direct formula 
for the Kostka matrix in terms of the Littlewood-Richardson 
coefficients \cite{macdonald:1979a}. 
The Kostka matrix is defined as the \emph{transition matrix} $M$ from monomial 
functions to Schur functions $K=M(s,m)$. Since $(m_\lambda)$ and $(h_\lambda)$ 
are dual \textit{wrt} the Schur scalar product, and the Schur functions are
self-dual, one obtains the transition matrix $K^*=M(s,h)$ for the basis change 
from complete symmetric functions to Schur functions. Noting that 
(\emph{dot} is the outer product here) $h_n = s_{(n)}$, $h_{\lambda} 
= h_{\lambda_1}\cdot\ldots\cdot h_{\lambda_r}$,
we compute (Sweedler indices as superscripts; note that
$(\lambda_i)$ is a one part partition and not a Sweedler index)
\begin{align}
K^*(s,h)_{\mu\lambda}&=(s_\mu\mid h_\lambda)\nn 
&= (s_\mu \mid s_{(\lambda_1)}\cdot \ldots \cdot s_{(\lambda_r)}) \nn
&= \left(\Delta^{(r-1)}(s_{\mu}) \mid s_{(\lambda_1)}\otimes\ldots\otimes
  s_{(\lambda_r)}\right)\nn
&= \sum (s_\mu^{(1)}\mid s_{(\lambda_1)})\ldots
              (s_\mu^{(r)}\mid s_{(\lambda_r)}) \nn
&= \sum_{\alpha_1,\ldots,\alpha_l}
C^\mu_{(\lambda_1)\alpha_1}\,C^{\alpha_1}_{(\lambda_2)\alpha_2}\,\ldots
C^{\alpha_{l-2}}_{(\lambda_{l-1})\alpha_l} \nn
&\equiv \sum_{\mu}\prod_i  (s_\mu^{(i)}\mid s_{(\lambda_i)})
\end{align}
where the Littlewood-Richardson coefficients emerge from the coproduct,
and $C^{\mu}_{\lambda(0)} = \delta^{\mu}_{\lambda}$ has been used.

Of course, similar calculations are possible for $M(h,s)$, $M(e,s)$, 
$M(s,e)$, \textit{etc}. From Macdonald, \textit{loc cit} (6.3)(3) one concludes further that if
$(u^\prime)$, $(v^\prime)$ are bases dual to $(u)$, $(v)$ then
\begin{align}
M(u^\prime,v^\prime) &= M(u,v)^\prime \,=\, M(u,v)^*
\end{align}  
holds, which shows that $K^*(s,h)=K(s,m)$ since $s^\prime=s$. 
\mye \noindent
Note that the above expansion can be used to compute the scalar product of the
monomial and the complete symmetric functions in the following way,
\begin{align}
M(m,m)^*_{\lambda,\mu}&= M(h,h)_{\lambda\mu} \,=\, (h_\lambda\mid h_\mu) \nn
&= \sum\sum\prod\prod \left( s^{(j)}_{(\lambda_i)}\mid s^{(i)}_{(\mu_j)}\right).
\end{align}
The double sum and product is reminiscent of the fact that we have to expand
both sides of the original scalar product. Especially interesting is the fact that 
\myb{Lemma}
\label{lemma:macdonaldkostkaprops}
\begin{itemize}
\item[(i)]
$M(e,m)_{\lambda\mu}=\sum_{\nu}K_{\nu\lambda}K_{\nu^\prime\mu}$ is the number
  of matrices of $0$'s and $1$'s with row sums $\lambda_i$ and column 
  sums $\mu_j$.
\item[(ii)]
$M(h,m)_{\lambda\mu}=\sum_{\nu}K_{\nu\lambda}K_{\nu\mu}$ is the number
  of matrices of non-negative integers with row sums $\lambda_i$ and 
column 
  sums $\mu_j$.
\end{itemize} \mye 
\noindent
{\bf Proof:} Macdonald \textit{loc cit}, (6.6)(i) and (ii) \mye \noindent
These matrices appear in a crucial way in renormalization formulae, 
supporting our claim, stated in the introduction, that symmetric 
functions may tell us valuable details about quantum field theory and 
renormalization (see concluding remarks below). 
The Hopf algebraic expansion is done by using the fact that one can 
introduce a resolution of the identity, 
\begin{align}
M(e,m) &= M(e,s)M(s,m) \,=\, M(s,m)M(h^\prime,s) , \nn
M(h,m) &= M(h,s)M(s,m) \,=\, M(s,m)M(h,s) ,
\end{align}
and using the above expansions.

\section{Basic Hopf algebra cohomology}

In this section we want to exploit the basic facts about Hopf algebra
cohomology as developed by Sweedler \cite{sweedler:1968a}. Firstly we notice
the well known theorem that

\myb{Theorem} Every coassociative coalgebra map $\Delta$ induces face
operators $\partial_n^i$ and coboundary maps $\partial_n$ from 
$H^{\otimes^n}$ to $H^{\otimes^{n+1}}$ as follows
\begin{align}
\partial_n^i &: H^{\otimes^n} \rightarrow H^{\otimes^{n+1}},
\qquad
\partial_n^i = 1\otimes \ldots \otimes \Delta \otimes
                \ldots \otimes 1\qquad \text{$i$-th place} \nn
\partial_n &: H^{\otimes^n} \rightarrow H^{\otimes^{n+1}},
\qquad
\partial_n = \sum_{i} \partial^i_n\,=\,
               \sum_{i} (-1)^i\,1\otimes \ldots \otimes \Delta \otimes
                \ldots \otimes 1\qquad \text{$i$-th place} \nn
\text{with}\quad&
\partial : H^\otimes \rightarrow H^\otimes,\qquad
\partial := \sum_n \partial_n,\qquad
\partial_{n+1}\circ\partial_n =0
\end{align}
\mye \noindent
{\bf Proof:} An inspection of the face maps $\partial_n^i$ shows that 
due to coassociativity, the coboundary maps $\partial_n$ obey the desired 
relation, $\partial_{n+1}\circ\partial_n=0$.
\mye \noindent
In fact, we will not use this setting, but one pulled down from $\lambda$-ring
addition to Hopf algebra convolution. Let $c_n \in \hom(H^{\otimes^n},\openZ)$ 
be a normalized unital $n$-linear form, which we call $n$-cochain. Unitality
is the property that $c(x_1\otimes \ldots \otimes 1 \otimes \ldots \otimes 
x_n)= e$ for any occurrence of a unit. Since the $\lambda$-ring addition is 
given by the convolution -- see appendix -- we will write the cohomology 
multiplicatively with respect to this outer convolution $(M,\Delta)$.

\myb{Definition} The convolution product of two $n$-cochains 
$c_n$, $c_n^\prime$ is given as
\begin{align}
(c_n*c_n^\prime)(x^1\otimes \ldots \otimes x^n) &=
c_n(x^1_{(1)}\otimes\ldots\otimes x^n_{(1)}) 
c_n^\prime(x^1_{(2)}\otimes\ldots\otimes x^n_{(2)})
\end{align}
which is Abelian, associative. Define $c_n*c_m=0$ if $n\not=m$.
A $n$-cochain is assumed to be normalized $c_n(1\otimes \ldots \otimes 1)=1$,
and hence known to be invertible $c_n^{-1}*c_n=\epsilon^{\otimes^n}=e$.
\mye \noindent
{\bf Aside:} The inverse was given by Milnor and Moore for general
homomorphisms under convolution \cite{milnor:moore:1965a} and rediscovered 
recently. The inverse of the identity map $\Id : H^\otimes \rightarrow
H^\otimes$ is the antipode and the recursive formula of Milnor and Moore 
reduced to this case is called `Connes-Kreimer' antipode formula in QFT. 
Since we will be able to use closed formulae for inverses, we need not take 
recourse in a computationally inefficient recursive definition.
\mye \noindent
We may now follow Sweedler and the development in
\cite{brouder:fauser:frabetti:oeckl:2002a} and define the coboundary 
operator acting on $n$-cochains in a multiplicative way \textit{wrt} convolution.
See appendix for the relation to $\lambda$-ring addition.
\myb{Definition}
\begin{align}
\partial_n^0 c_{n-1}(x^1\otimes \ldots \otimes x^n) &:=
 \epsilon(x^1)c_{n-1}(x^2\otimes \ldots \otimes x^n) \nn  
\partial_n^i c_{n-1}(x^1\otimes \ldots \otimes x^n) &:=
 c_{n-1}(x^1\otimes x^ix^{i+1}\ldots \otimes x^n) \nn  
\partial_n^n c_{n-1}(x^1\otimes \ldots \otimes x^n) &:=
 c_{n-1}(x^1\otimes \ldots \otimes x^{n-1}) \epsilon(x^n) \nn
\partial_n c_n &:= \partial_n^0 c_n * \partial_n^1 c_n^{-1} *
      \partial_n^2 c_n \ldots * \partial_n^n c_n^{\pm 1} \nn
   \partial_{n+1}\circ\partial_n &= e \qquad\text{with}\qquad
e:= \epsilon\otimes \ldots\otimes \epsilon.
\end{align}
We will denote the coboundary map simply as $\partial$ if the context is 
clear.
\mye \noindent
Note, that since we wrote the cohomology multiplicatively, the trivial element
is the identity $e=\epsilon^\otimes$, the $n$-fold tensor product of the
counit. Furthermore, the alternating sum of face-maps $\sum \pm \partial_n^i$
translates into a `see-saw' product of maps and inverse maps. We can use 
the coboundary map to classify $n$-cochains as follows:
\myb{Definition}
A $n$-coboundary $b_n\in B^n$ is a $n$-cochain fulfilling $b_n = 
\partial c_{n-1}$. A $n$-cocycle $c_n \in Z^N$ is a $n$-cochain fulfilling 
$\partial c_n=e$. An $n$-cochain which is neither a coboundary nor a cocycle 
will be called generic. 
\mye \noindent
It follows from the definition, that an $n$-coboundary is a
$n$-cocycle. Furthermore, the $n$-cocycles form an Abelian group under
convolution. Hence one may build the quotient of $n$-cocycles by 
$n$-coboundaries to form the $n$-th cohomology group, $H^n = {Z^n}/{B^n}$
Indeed, this allows one to compute \textit{eg} the `Betti' numbers of the complex 
(see \cite{lefschetz:1975a}). From \cite{brouder:fauser:frabetti:oeckl:2002a} we may take the
characterization of 1-cocycles to be
\myb{Theorem} A 1-cochain is a 1-cocycle if and only if it is an algebra
homomorphism, 
\begin{align}
\label{eq:cocycle}
e\,=\,\partial c_1(s_\lambda\otimes s_\mu) &=
\sum_{\alpha}\sum_{\beta} c_1(s_{\alpha})c_1(s_{\beta})
c_1^{-1}(s_{\lambda/\alpha}\, s_{\mu/\beta}) 
\,=\, \epsilon(s_\lambda)\epsilon(s_\mu)\nn
\Leftrightarrow \quad c_1(s_{\lambda}\, s_{\mu}) &=
  c_1(s_{\lambda}) c_1(s_{\mu})
\end{align}
\mye \noindent
where we have used the outer Hopf algebra in the convolution and exemplified
the condition on a Schur function basis. This basic fact will suffice to make
some observations and classifications in the sequel related to the 
Schur function series introduced by Littlewood.

\subsection{Littlewood-King-Wybourne infinite series of Schur functions}

Littlewood \cite{littlewood:1940a} gave a set of Schur function series, 
which allowed him to formulate various identities in an extremely compact 
notation. These identities have been extended by King, Dehuai and Wybourne
\cite{king:dehuai:wybourne:1981a} and later by Yang and Wybourne
\cite{yang:wybourne:1986a}; we follow the presentation of the latter.

An $S$-function series is an infinite formal sum of Schur functions given via a
generating function. It turns out, that the most basic Schur function series
is the so called $L$-series, 
\begin{align}
L &= \prod_{i=1}^\infty (1-x_i)
\end{align}
from which the others may be derived. It is possible to give the $S$-function
content of the series
\begin{align}
L &= \sum_{m=0}^\infty (-1)^{m} s_{(1^{m})} = 
     \sum_{m=0}^\infty (-1)^{m} \{ 1^{m} \} 
\end{align}
where we have introduced the common notation $\{ \lambda\}$ of Littlewood
for a Schur function $s_\lambda$. Furthermore it is convenient to follow
Yang and Wybourne to introduce the conjugate (with respect to transposed 
partitions) series and the inverse conjugate series as
\begin{align}
L^\dagger &= (\widetilde{L})^{-1} \,=\, \widetilde{L^{-1}} \nn
          &= \prod_{i=1}^\infty (1+x_i)^{-1}\,=\, \sum_{m=0}^\infty 
          (-1)^{m} \{ m\}.
\end{align}
Notice that taking the conjugate is equivalent to the transformation $x_i
\rightarrow -x_i$, which can be viewed as a plethysm (written with the
concatenation symbol $\circ$ as before),
\begin{align}
L^\dagger &= L(-x_i) \,=\, (-\{1\})\circ L.
\end{align}
For a Hopf algebraic approach to plethysm see \cite{scharf:thibon:1994a}.
The other series are then derived in a similar manner, see
\cite{yang:wybourne:1986a}. $S$-function series come in pairs which are
mutually inverse and consecutively named. One finds
\begin{align}
AB=1,\quad CD=1,\quad EF=1\ldots, \quad LM=1,\quad PQ=1,\ldots \quad VW=1,\ldots \,
\end{align}
which may be arranged as in Table 2, following \cite{yang:wybourne:1986a}.
\begin{table}[h]\begin{center}
\caption{$S$-function series: Type, Name=Product, Schur function content, and
plethysm.}
\begin{align}
\label{eq:series}
\begin{array}{c@{~~~}c@{\,=\,}l@{~~}l@{~~}l@{~~}l}
\hline\hline\\
L      & L & \prod_i(1-x_i) & \sum_m(-1)^m\{1^m\} & L(x_i) 
& \{1\}\circ L \\
L^{-1} & M & \prod_i(1-x_i)^{-1}  & \sum_m\{m\} & L(x_i)^{-1} 
& \{1\}\circ L^{-1}\\
\tilde{L} & P & \prod_i(1+x_i)^{-1}  & \sum_m (-1)^m\{m\} & L(-x_i)^{-1} 
& (-\{1\})\circ L\\
L\dagger & Q & \prod_i(1+x_i) & \sum_m \{1^m\} & L(-x_i) 
& (-\{1\})\circ L^{-1}\\
A & A & \prod_{i<j}(1-x_ix_j) & \sum_\alpha (-1)^{\omega_\alpha/2}
\{\alpha\} & L(x_ix_j) (i<j) 
& \{1^2\}\circ L\\
A^{-1} & B & \prod_{i<j}(1-x_ix_j)^{-1} & \sum_\beta \{\beta\} & 
L(x_ix_j)^{-1} (i<j) 
& \{1^2\}\circ L^{-1}\\
\tilde{A} & C & \prod_{i\le j}(1-x_ix_j) & \sum_\gamma 
(-1)^{\omega_\gamma/2}\{\gamma\} & L(x_ix_j) (i \le j) 
& \{2\}\circ L\\
A^\dagger & D & \prod_{i\le j}(1-x_ix_j)^{-1} & 
\sum_\delta \{\delta\} & L(x_ix_j)^{-1} (i \le j)  
& \{2\}\circ L^{-1}\\
V=\tilde{V} & V & \prod_{i}(1-x_i^2) & 
\sum_{p,q} (-1)^p\{\widetilde{p+2q},p\} & L(x_i^2)  
& (\{2\}-\{1^2\}) \circ L\\
V^{-1}=V^\dagger & W & \prod_{i}(1-x_i^2)^{-1} & 
\sum_{p,q} (-1)^p \{ p+2q,p \} & L(x_i^2)^{-1}  
& (\{2\}-\{1^2\})\circ L^{-1}\\[1ex]
\hline
\end{array} \nonumber
\end{align}
\end{center}\end{table}
The remaining series are $E=LA$, $F=L^{-1}A^{-1}$, $G=L^\dagger A$,
$H=\tilde{L}A^{-1}$, $R=L\tilde{L}$ and $S=L^{-1}L^\dagger$. The 
partitions $\{\alpha\}$ associated with the $A$ series are defined as follows (in 
Frobenius notation)
\begin{align}
(\alpha) &= (a_{1} a_{2} \ldots a_{r} \mid  a_1\! +\! 1   a_2 \! +\! 1  \ldots  
 a_r \! +\! 1 )
\end{align}
and for the $B$, $C$ and $D$ series, $\{\beta\}$ is the transpose of
$\{\alpha\}$, $\{\delta\}$ has only even parts and $\{\gamma\}$ is its
transpose.  $\{\epsilon\}$ (not in the table, but related to $E$) has
only self conjugate partitions and $\{\zeta\}$ (not in the table, but
related to $F$) contains all partitions.  Finally, $\omega_\lambda$ indicates
the weight of the partition under consideration.

\subsection{Cochain induced branching operators and series}

The series in the above subsection play a fundamental role in the theory of
group characters. We will show now, in which way these series are
related with a special class of endomorphisms acting on $\Lambda$. Those
endomorphisms play also a considerable role in QFT, where they are related to
time, operator, and normal ordering of quantum fields 
\cite{brouder:fauser:frabetti:oeckl:2002a}.

In the following, we are interested in those endomorphisms of the ring of
symmetric functions, which can be derived from 1-cochains. We introduce 
lower case symbols $\phi$ for 1-cochains, and the related operators 
$/\Phi : \Lambda \rightarrow \Lambda$, denoted by upper case letters, in the 
following way (\textit{cf} (\ref{eq:dualaction}) above):
\myb{Definition} An invertible branching operator is an endomorphism $/\Phi$
based on a 1-cochain $\phi$ via
\begin{align}
\label{eq:bo}
/\Phi(s_{\lambda}) &:= (\phi\otimes \Id)\circ\Delta(s_\lambda)
\,=\, \sum_\alpha \phi(s_\alpha) s_{\lambda/\alpha}
\end{align}
such that $\phi(s_\mu) \in \openZ$.
\mye \noindent
We will now use the
above-displayed results on cohomology to characterize the resulting maps.

\subsubsection{Skewing by a series}

In group branching laws and formal $S$-function manipulations, one is
interested in computing the skews of a particular irreducible representation 
described by a partition $\lambda$, with the elements of the series under
consideration. Consider for example $M$,
\begin{align}
s_{\lambda}/M &= \sum_{m\in M} s_{\lambda/(m)} \,=\, s_{\lambda/(0)} 
+ s_{\lambda/(1)} + s_{\lambda/(2)} + \ldots ,
\end{align}
where the resulting set of terms is actually finite, since 
$s_{\lambda/\mu}$ is zero if the weight of $\mu$ is greater than the weight 
of $\lambda$. In view of the outer coproduct,
\begin{align}
\Delta(s_{\lambda}) &= s_{\lambda/(0)} \otimes s_{(0)}
+ s_{\lambda/(1)} \otimes s_{(1)} + s_{\lambda/(1^2)} \otimes s_{(1^2)}
+ s_{\lambda/(2)} \otimes s_{(2)} + s_{\lambda/O(3)} \otimes s_{O(3)},
\end{align}
(where $O(3)$ means terms of weight equal or higher than 3), it is 
clear that a 1-cochain (linear form) can be defined to act on one tensor
factor such that the resulting terms form the $/M$ skew series of
$s_{\lambda}$. Generically for an arbitrary series $\Phi$ 
the 1-cochain $\phi$ is the corresponding `characteristic function',
\begin{align}
\phi_{\lambda} \equiv \phi(s_{\lambda}) &= \left\{\begin{array}{cl}
1 & \text{if $\lambda \in \Phi$ } \\
0 & \text{otherwise.}
\end{array}\right.
\end{align}
This motivates \textit{a posteriori} the name branching operator 
given to the above defined operators $/\Phi$, and henceforth we
adopt the notation $\Phi\cong \,/\Phi$ (see below). 

\myb{Lemma} The inverse branching operator is 
\begin{align}
\Phi^{-1} &= (\phi^{-1}\otimes \Id)\circ \Delta
\end{align}
where the inverse of the 1-cochain $\phi$ is with respect to the outer convolution.
\mye \noindent
The outer product is trivial here since the value of the 1-cochain is in 
$\openZ$.\\
{\bf Proof:} We compute the composition of the two operators directly using
coassociativity of the outer coproduct
\begin{align}
\Phi^{-1}(\Phi(s_\lambda)) &= \sum_{\alpha,\beta}
\phi^{-1}(s_{\alpha})\,\phi(s_{\beta/\alpha}) s_{\lambda/\beta} 
\,=\, \sum_{\beta} \epsilon(s_{\beta}) s_{\lambda/\beta}
\,=\, s_{\lambda}
\end{align}
\mye \noindent
Thus the obvious inverse operation, skewing by the inverse series 
(with respect to the outer product), has an internal structure 
governed by outer convolution at the level of the underlying 1-cochain.
Finally this allows us to form the following new product $\outerM_\phi$, which will be set 
in a more general context in the next section. 
\myb{Definition} The $\phi$-deformed outer product $\outerM_\phi$ is defined as
\begin{align}
\outerM_\phi(f\otimes g) &= f\circ_{\phi} g = \Phi^{-1}(\outerM(\Phi(f)\otimes \Phi(g)))
\end{align}
\mye \noindent

\subsubsection{Classifying branching operators}

From cohomology we know already that there are cochains of different
types, for example 1-coboundaries, 1-cocycles and generic 1-cochains.  Hence we
expect that this difference shows up in the nature of the branching
operators induced by these 1-cochains.

\paragraph{Trivial 1-cochain:} The trivial 1-cochain is the counit. By
definition, the counit acts such that the coproduct action is void, and we get
as branching operator for the trivial 1-cochain, the identity.

\paragraph{1-coboundaries:} We have shown above that the set of 
1-coboundaries is empty. Hence we have no operators from this class.

\paragraph{1-cocycles:} From the cohomology we learned that the
property of being a 1-cocycle $\phi$ is equivalent to being an algebra 
homomorphism, $\phi(\{\lambda\}\cdot \{\mu\}) = \phi(\{\lambda\})
\phi(\{\mu\})$, alternatively written as $(s_\lambda \cdot s_\mu)/\Phi
= s_\lambda/\Phi\cdot s_\mu/\Phi$. The same is true for the inverse 
1-cocycle $\phi^{-1}$. This allows us to state that
\myb{Lemma} The outer product $\outerM$ and the $\phi$-deformed outer product 
$\outerM_\phi$ are isomorphic if and only if $\phi$ is a 1-cocycle. 
\mye \noindent
{\bf Proof:} This follows directly from the definition, and the fact that the
inverse branching operator is an algebra homomorphism too.
\mye \noindent
Hence we find 
\begin{align}
\text{$\phi$ is a 1-cocycle} \quad&\Leftrightarrow\quad
\phi \in \text{alg-hom}((\Lambda^\otimes,M),(\Lambda^\otimes,M)).
\end{align}
As we will see later, these deformed products are far from being empty
constructs.  Indeed, we have a coalgebra action and an augmentation,
the counit $\epsilon$, and one checks easily that 
\myb{Lemma} The
augmented outer (comodule) algebras $(\Lambda^\otimes,\outerM,\epsilon)$ and
$(\Lambda^\otimes,\outerM_\phi,\epsilon)$ are non-isomorphic.  \mye \noindent 
{\bf Proof:} Since the counit is \emph{not} transformed, it acts
differently on the two algebras. \mye \noindent
Indeed the \textit{isomorphic} structures are related by adopting as 
the $\phi$-deformed counit
\begin{align}
\epsilon &\rightarrow \epsilon\circ\Phi\,=\,\phi*\epsilon \, \equiv \, \phi,
\end{align}
which is in general different from $\epsilon$. This may be worth exemplification, so for a general $\Phi$ we 
compute the action of $\epsilon$ on $s_{(1)}\cdot s_{(1)}$ and 
$s_{(1)}\circ_{\phi} s_{(1)}$: 
\begin{align}
\label{eq:3-69}
a)&\quad \epsilon(s_{(1)}\cdot s_{(1)}) \,=\, 
         \epsilon(s_{(2)}+s_{(1^2)}) \,=\,0\nn
b)&\quad \epsilon(\Phi^{-1}(\Phi(s_{(1)}) \cdot \Phi(s_{(1)})))
 \,=\, \epsilon\circ\Phi^{-1}\big( (s_{(1)}\phi_{0}+ s_{0}\phi_{(1)}) 
\cdot (s_{(1)}\phi_{0}+ s_{0}\phi_{(1)}) \big) \nn
&\qquad\,=\,
\epsilon\circ \Phi^{-1}(s_{(1)}\cdot s_{(1)} + 2\phi_{(1)} s_{(1)} + {\phi_{(1)}}^2 )
 \nn &\qquad \,=\,  {\phi_{(1^{2})}}^{-1} + {\phi_{(2)}}^{-1} + {\phi_{(1)}}^2 
\end{align}
which is $ \not=0$ in general.  (In the last line the
normalisation ${\phi_{0}}^{-1}=\phi_{0}=1$ and the definition of
$\epsilon$ has been used).

\paragraph{Generic 1-cochain:} For generic 1-cochains the above consideration
fails, hence we can only state that a deformed outer product $\outerM_\phi$ based 
on a generic 1-cochain is non-isomorphic to the original outer product. The 
relation between these two products will become clear in the next section.

\subsubsection{Classifying series}

In summary, the cohomological properties of the 1-cochains classify
the associated branching operators $\Phi$.  Besides the identity, we
get those based on 1-cocycles which produce isomorphic products, and
generic ones.  Since we \textit{defined} the $S$-function series using
branching operators, we can in turn classify the series into 3
families according to the underlying cochains: those based on
coboundaries (empty for 1-cochains), those based on cocycles and those
based on generic cochains.

\myb{Theorem} The series derived from branching operators based on 
1-cocycles $\phi$ fulfil
\begin{align}
\label{eq:3-70}
\{\lambda\cdot \mu\}/\Phi &= \{\lambda\}/\Phi \cdot \{\mu\}/\Phi 
\end{align}
reflecting the homomorphism property.
\mye \noindent
{\bf Proof:} Note that $/\Phi$ was defined as $(\phi\otimes \Id)\circ 
\Delta$. Using the product-coproduct homomorphism axiom of bialgebras we
compute
\begin{align}
(\phi\otimes\Id)(\Delta\circ M) &= 
(\phi\otimes\phi\otimes M)(\Id  \otimes {\sf sw}\otimes \Id  )(\Delta\otimes\Delta)
\end{align} 
which, after reformulating using $/\Phi$ yields the assertion.
\mye \noindent
It is now a matter of explicit calculation to check which series 
fulfil the cocycle property. 
The generic case (for the remaining series) comes up in two different forms depending on the appearance of 
the antipode. One finds
\myb{Theorem}
\label{th:branchop} 
\begin{itemize}
    \item[(i)] Branching operators based on the series $L,M,P,Q,R,S,V,W$ 
    fulfil  (\ref{eq:3-70}).
    \item[(ii)] Branching operators based on generic 1-cochains
of the $B,D,F,H$ series satisfy
\begin{align}
\label{eq:NL}
 \quad\{\lambda\cdot \mu\}/\Phi &= \sum_{\zeta} 
\{\lambda\}/(\zeta\cdot\Phi) \cdot \{\mu\}/(\zeta\cdot\Phi).
\end{align}
     \item[(iii)] Branching operators based on generic 1-cochains
of the $A,C,E,G$ series satisfy
\begin{align}
\label{eq:NL2}
 \quad\{\lambda\cdot \mu\}/\Phi &= 
\sum_{\zeta} \{\lambda\}/(\zeta\cdot\Phi) \cdot 
\{\mu\}/(\antip(\zeta)\cdot\Phi)\nn
&= \sum_{\zeta} (-1)^{\omega_\zeta}\,
\{\lambda\}/(\zeta\cdot\Phi) \cdot \{\mu\}/(\zeta^\prime\cdot\Phi).
\end{align}
where the summation is over \textit{all} Schur functions $\zeta$ (\textit{ie}
using the $F$ series).  
\end{itemize}\mye \noindent
{\bf Proof:} The proof and further material in this direction can be found in
\cite{king:dehuai:wybourne:1981a,black:king:wybourne:1983a,%
king:wybourne:yang:1989a}. A direct proof is based on formal 
evaluation of the outer coproduct, based on 
(\ref{eq:theta:co:xytrick}). For the series $D$ (Table 2) we have for 
example
\begin{align}
    D(x,y) &= \prod_{i \le j}(1-x_{i}x_{j})^{-1}\prod_{\ell \le 
    m}(1-y_{k}y_{\ell})^{-1} \prod_{k ,n }(1-x_{k}y_{n})^{-1} \nn
    &= D(x)D(y)\sum_{\zeta}s_{\zeta}(x)s_{\zeta}(y) \nn
    &= \sum_{\zeta} D(x) s_{\zeta}(x) \cdot D(y)s_{\zeta}(y) \nonumber
\end{align}
where the fundamental Cauchy identity has been used in the second 
line \cite{macdonald:1979a}.
\mye \noindent
We shall call the series $L,M,P,Q,R,S,V,W$ \emph{group like}, because of 
the above outer coproduct property and in view of (\ref{eq:3-70}) (see 
below). In fact, it is easy to see that the above argument goes 
through for \textit{any} series defined by a generating function of 
the form $\Phi = \prod_{i}(1-f(x_{i}))^{s}$ for some polynomial $f(x)$.
Note finally that the second two lists are made from mutually
inverse elements $A,B$; $C,D$; $E,F$; and $G,H$.

\subsection{Products versus orderings}

It is worth noting at this stage that a structural analogy with
quantum physics can be seen in relation to products and operator
ordering.  Just as in QFT both normal and time ordered products are
needed, so too in symmetric function theory different products induced
by branching operators emerge: the rings $(\Lambda, \outerM)$ and
$(\Lambda, \outerM_{\phi})$ related by a cocycle are $\phi$-isomorphic
(but differ under the augmentation by the co-unit, $\epsilon$).  This
analogy is elaborated in the concluding remarks below in terms of an
\textit{active} versus \textit{passive} analysis of the role of the
branching endomorphisms.  In order to derive further Hopf-algebraic
motivated insights into these formulae, we need to consider
2-cochains, and reconsider the Schur scalar product and its
convolutive inverse in a more elaborated approach to deformed
products, namely that of cliffordization.

\section{Cliffordization}

\subsection{Definition}

The first occurrence of a cliffordization is to our knowledge in
Sweedler \cite{sweedler:1968a}, where it was derived from a smash product in a
cohomological context.  Later, Drinfeld \cite{drinfeld:1987a} invented the twisting of
products, also induced by a Hopf algebraic construction related to a
smash product.  The term `Cliffordization' was
coined in \cite{rota:stein:1994a,rota:stein:1994b}.  There a coalgebra
structure was also employed, but not in general a Hopf algebra, and the
combinatorial aspects were emphasized.  Since we follow the
direction taken by Rota in the treatment of symmetric functions, it 
seems reasonable to stick to this technical term.  A more Hopf-algebraic
motivated approach is included in
\cite{brouder:fauser:frabetti:oeckl:2002a}.

Let us assume that we start with a Hopf algebra, but focus our interest
for the moment on the algebraic part of it.  Given a linear space
underlying an algebra, we are interested to induce a family of
deformed products using a pairing.  Special properties of the pairing
will ensure special properties of the deformed product.  In general,
such a pairing should be a measuring, to keep the Hopf algebraic
character of the whole structure \cite{schmitt:1999a}.  However, we
will drop this requirement.  The resulting structure is a comodule
algebra, hence an algebra with a coaction, which is not necessarily a
bialgebra or Hopf.

\myb{Definition} A self pairing is a linear map $\pi : \Lambda^\otimes \otimes 
\Lambda^\otimes \rightarrow \Lambda^\otimes$. A scalar pairing has $\openZ$ as
codomain.
\mye \noindent
We will be interested in scalar self pairings. Indeed, we have already made use of
the `natural' pairing of symmetric functions, that is, the Schur scalar
product. This allows us to give the following
\myb{Definition} A cliffordization is a deformation of a comodule, or
possibly a Hopf, algebra $(\Lambda,\cdot)$ into a twisted comodule
algebra on the same space $\Lambda$ equipped with the circle product
\begin{align}
x\circ y &= \sum \pi(x_{(1)}\otimes y_{(1)}) \, x_{(2)}\cdot y_{(2)} 
\nonumber
\end{align}
\mye \noindent
The name cliffordization\footnote{The `circle product' is \emph{not} to be confused with plethysm, which will
play no role for the moment.} stems from the fact that if $\pi$ is a 'scalar product' and
$(V^\wedge,\wedge)$ a Grassmann algebra, then $\circ$ is the endomorphic
Clifford product induced by $\pi$ \cite{fauser:2002c,fauser:2002d}.

In the case of Schur functions a twisted or cliffordized product can
be given by the Schur scalar product \emph{or} its inverse playing the
role of the pairing $\pi$.  From Hopf algebra theory we know further,
that the inverse of a 2-cocycle, \textit{wrt} the convolution, is given by
acting with the antipode in the first or second argument,
\begin{align}
\pi(x \otimes y) &= \pi( {\sf S}(x)\otimes {\sf S}(y)) \nn
\pi(x \otimes y)^{-1} &= 
\pi( {\sf S}(x)\otimes y) \,=\, \pi( x\otimes {\sf S}(y))
\end{align}
which allows the introduction of the (off diagonal) convolutive 
\emph{inverse} Schur scalar product as
\begin{align}
(s_{\lambda}\mid s_{\mu})^{-1} &= (-1)^{\omega_{\mu}}\,
(s_{\lambda}\mid s_{\mu^\prime}).
\end{align}
Thus the scalar product cliffordizations read
\begin{align}
\label{eq:cliff}
s_{\lambda}\circ s_{\mu} &= \sum_{\alpha,\beta}
(s_{\alpha}\mid s_{\beta}) \, s_{\lambda/\alpha}\cdot s_{\mu/\beta} \nn
&= \sum_{\alpha}s_{\lambda/\alpha}\cdot s_{\mu/\alpha},\nn
s_{\lambda}\circ_{\sf S} s_{\mu} &= \sum_{\alpha,\beta} (-1)^{\omega_{\beta}}\,
(s_{\alpha}\mid {\sf S}(s_{\beta})) \, 
s_{\lambda/\alpha}\cdot s_{\mu/\beta} \nn
&= \sum_{\alpha} (-1)^{\omega_{\alpha}} 
s_{\lambda/\alpha}\cdot s_{\mu/\alpha^\prime},
\end{align}
since the Schur functions are orthogonal. 
Clearly the two formulae in (\ref{eq:cliff}) and those
in (\ref{eq:NL}), (\ref{eq:NL2}) are based on the Schur scalar product and its
inverse, up to the additional appearance of a series. The situation 
is summarised in the following section.

\subsection{Classifying 2-cochains and cliffordization}

Scalar pairings can be regarded as 2-cochains. It is therefore
convenient to classify them in analogy with the 1-cochains. We find the
following:
\paragraph{Trivial 2-cochain:} The trivial 2-cochain is the map $e^{\otimes^2}=
\epsilon\otimes \epsilon$. Hence in substituting $\pi=e^{\otimes^2}$ in
(\ref{eq:cliff}) one sees that the product just remains unaltered, and this
yields the identity deformation.
\paragraph{2-coboundaries:} A 2-coboundary is a pairing which is derived from
a 1-cochain via the coboundary operator, $\pi = \partial \phi$ for a 
1-cochain $c_{1} \equiv \phi$. Looking at
(\ref{eq:cocycle}) we see that these 2-cochains are of the form
\begin{align}
    \label{eq:2cob}
\pi_{\phi}(x\otimes y) &= \partial \phi(x\otimes y) \,=\,
\phi(x_{(1)}) \phi(y_{(1)}) \phi^{-1}(x_{(2)}y_{(2)})\,.
\end{align}
This shows, that the \emph{group like} deformations by a series 
can be be equally well addressed as a cliffordization by a 2-coboundary.
In fact, one is able to rearrange the $\phi$-deformed outer product 
in the form
\begin{align}
\outerM_\phi(x\otimes y) &= \Phi^{-1}(\outerM(\Phi(x)\otimes \Phi(y))
\,\equiv  \sum \pi_\phi(x_{(1)}\otimes y_{(1)}) x_{(2)}\cdot y_{(2)}
\end{align}
\textit{if and only if} the 2-cochain $\pi$ is a coboundary \cite{fauser:2002c}.
\paragraph{2-cocycles:} A 2-cocycle is characterized by $\partial \pi = e$. 
It is well known from Hopf algebra theory, and from the theory of $*$-product
deformations, that deformations induced by a 2-cocycle yield associative
products (for details see \cite{brouder:fauser:frabetti:oeckl:2002a} and references
therein). Moreover, since our 2-cochains are assumed to be
normalized, they are invertible. One finds
\begin{align}
  (\circ_\phi)_{\phi^{-1}} &= \circ_{\phi*\phi^{-1}} \,=\, \circ_{e} 
\end{align}
showing the invertibility of the deformation process. For the case of symmetric functions 
and the Schur scalar product and its inverse, the cliffordizations 
are (\ref{eq:cliff}), as already noted above.  
\paragraph{Generic 2-cochains:} The deformed circle product based on a generic
2-cochain cannot be associative and we will here not consider such
deformations. 
\paragraph{Mixed 2-cocycles and 2-coboundaries:} It is possible to draw the
analogy that 2-coboundaries are topological `gauges', while the
2-cocycles are generic topological `fields'. Since the space of cocycles,
boundary or not, forms a group, we can pick a section in the orbits of
2-coboundaries, \textit{ie} gauge the 2-cocycles by adding a 2-coboundary as
\begin{align}
\pi &= c_2 * \partial c_1.
\end{align}
Taking one of the formulas from (\ref{eq:cliff}) and a 2-coboundary which we
know to be derived from a 1-cochain, that is from a series, we have 
\begin{align}
\label{eq:4-83}
x \circ_\phi y &= \sum ((~.~\mid ~.~)*\partial_\phi)(x_{(1)}\otimes y_{(1)})\,
x_{(2)}\cdot y_{(2)}\nn
x \circ_{{\sf S},\phi} y &= 
\sum ((~.~\mid {\sf S}(~.~))*\partial_\phi)(x_{(1)}\otimes y_{(1)})\,
x_{(2)}\cdot y_{(2)}
\end{align}
which can be rewritten in the form analogously to (\ref{eq:NL}), 
(\ref{eq:NL2}) (for the 
generic cases of Theorem 3.12),
\begin{align}
\label{eq:4-84}
(s_\lambda\cdot s_\mu)/\Phi &=\sum (s_\alpha\mid s_\beta) 
s_{\lambda/(\alpha\Phi)}\cdot s_{\mu/(\beta\Phi)}\nn
&= \sum s_{\lambda/(\alpha\Phi)}\cdot s_{\mu/(\alpha\Phi)}\nn
(s_\lambda\cdot s_\mu)/_{\sf S}\Phi &=\sum (s_\alpha\mid {\sf S}(s_\beta)) 
s_{\lambda/(\alpha\Phi)}\cdot s_{\mu/(\beta\Phi)}\nn
&= \sum (-1)^{\omega_{\alpha}}\,
s_{\lambda/(\alpha\Phi)}\cdot s_{\mu/(\alpha^\prime\Phi)}
\end{align}
where we have used (\ref{eq:2cob}) and $(s_{\lambda}/\alpha)/\Phi = s_{\lambda/(\alpha\Phi)}$.
Hence we proved the following two lemmas.  
\myb{Lemma} (\ref{eq:NL}), (\ref{eq:NL2}) are cliffordizations \textit{wrt}
the Schur scalar product or its inverse, in convolution with a
2-coboundary induced by an $S$-function series (or branching
operator).  These formulae are precisely the Newell-Littlewood 
products (see below). \mye \noindent 
\myb{Lemma} The two instances of the
formulae in (\ref{eq:NL}), (\ref{eq:NL2}) are mutually inverse to one another if the
involved series are mutually inverse, \textit{eg} $A,B$; $C,D$; \textit{etc}.  \mye
\noindent 
\textbf{Aside:} Before discussing the group aspects of this result, we might
dwell again on the striking structural analogy to quantum field theory
developed so far.  The `geometry' -- \textit{ie} the analogue of
quantization-- is induced by the Schur scalar product (or its
inverse), while the ordering structure or basis choice is maintained
by the 2-coboundaries, the `propagators' in the quantum field
theoretical language.  Further arguments
will strengthen this after the branching laws have been considered
(see also the concluding remarks below). 

\subsection{8 possible Cliffordizations}

To finish the technical parts of the discussion of cliffordization, we want to
present an overview on what kind of open possibilities remain to be
explored. Indeed, looking at the stock of 'natural' structures in symmetric
function theory, we find the Schur scalar product and its inverse, the outer
and inner products and the outer and inner coproducts. Examining the
definition of cliffordization, one notes that it involves two coproducts, one
2-cocycle and 1 product. If we consider the inverse Schur scalar product not
as essentially different from the Schur scalar product, then we find a total
of 8 different possibilities to employ the inner and outer products and
coproducts in cliffordization. We obtain the cliffordizations and the
grades for a products of homogenous elements as
\begin{align}
f\circ_1 g &= \sum \pi(f_{(1)}\otimes g_{(1)})\, \outerM(f_{(2)}\otimes g_{(2)}),
\qquad \vert n\vert\otimes \vert m\vert \rightarrow 
\oplus_{r} \vert n+m-2r\vert \nn
f\circ_2 g &= \sum \pi(f_{(1)}\otimes g_{(1)})\, \innerm(f_{(2)}\otimes g_{(2)}),
\qquad \vert n\vert\otimes \vert m\vert \rightarrow 
\oplus_{r}  \delta_{nm} \vert n-r\vert \nn
f\circ_3 g &= \sum \pi(f_{[1]}\otimes g_{(1)})\, \outerM(f_{[2]}\otimes g_{(2)}),
\qquad \vert n\vert\otimes \vert m\vert \rightarrow 
  \vert m\vert \nn
f\circ_4 g &= \sum \pi(f_{[1]}\otimes g_{(1)})\, \innerm(f_{[2]}\otimes g_{(2)}),
\qquad \vert n\vert\otimes \vert m\vert \rightarrow 
 \delta_{n,m-n} \vert n\vert \nn
f\circ_5 g &= \sum \pi(f_{(1)}\otimes g_{[1]})\, \outerM(f_{(2)}\otimes g_{[2]}),
\qquad \vert n\vert\otimes \vert m\vert \rightarrow 
 \vert n\vert \nn
f\circ_6 g &= \sum \pi(f_{(1)}\otimes g_{[1]})\, \innerm(f_{(2)}\otimes g_{[2]}),
\qquad \vert n\vert\otimes \vert m\vert \rightarrow 
 \delta_{n-m,m} \vert m\vert \nn
f\circ_7 g &= \sum \pi(f_{[1]}\otimes g_{[1]})\, \outerM(f_{[1]}\otimes g_{[1]}),
\qquad \vert n\vert\otimes \vert m\vert \rightarrow 
\delta_{nm} \vert 2n\vert \nn
f\circ_8 g &= \sum \pi(f_{[1]}\otimes g_{[1]})\, \innerm(f_{[1]}\otimes g_{[1]}),
\qquad \vert n\vert\otimes \vert m\vert \rightarrow 
 \delta_{nm} \vert n\vert
\end{align}
where we used the Brouder-Schmitt convention on coproducts and
$\outerM,\innerm$ for the outer and inner product of symmetric
functions.  The right column displays the grades obtained in
multiplying two homogenous elements of grade $\vert n\vert$ and $\vert
m\vert$.  Of course, any 2-cocycle induces in this way 8
cliffordizations.  This raises the question about a classification of
all 2-cocycles acting on the ring of symmetric functions
$\Lambda^{\otimes^2}$.  While we have considered the Schur scalar
product for the ring $\otimes_{\openZ}\Lambda$, the appearance of
Hall-Littlewood symmetric functions and Macdonald symmetric functions
show clearly that these cases are tied to ring extensions.  From
Scharf and Thibon's approch to inner plethysm
\cite{scharf:thibon:1994a} (p.  33) it is obvious, that the change of
one structure map in a convolution changes its properties
dramatically.  One finds
\begin{align}
\label{eq:4-87}
f\,+_{\!\Lambda}\, g &\quad\Leftrightarrow 
\outerM\circ(f\otimes g)\circ \Delta \nn
f\,\cdot_{\!\Lambda}\, g &\quad\Leftrightarrow 
\outerM\circ(f\otimes g)\circ \delta
\end{align}
Hence changing the coproduct from outer to inner changes the $\lambda$-ring
operation induced by convolution from addition to multiplication (see 
appendix). In fact,
this is the source of the difference between the outer and inner branching rules.
This observation gives a hint, as to the way in which the above cliffordizations may 
change if inner products and coproducts replace the outer products and 
coproducts. A few quite amazing properties can be easily derived, but we 
will not enter this subject here. 

\subsection{Branching rules:  
$U(n) \!\downarrow\!   O(n)$; $U(n) \!\downarrow\!   Sp(n)$;
$O(n) \!\uparrow\!   U(n)$; $Sp(n) \!\uparrow\! U(n)$ and product rules:
$Sp(n) \! \times \!Sp(n) \!\downarrow\!   Sp(n)$, $O(n) \! \times 
\! O(n) \!\downarrow\! O(n)$}

The above-discussed relations for Schur functions reflects their
interpretation as universal characters \cite{king:1990a}. We will not deal
with the problem of modification rules needed for actual evaluation of the
reduced group characters, but follow in our presentation King \textit{loc cit}. Our 
aim thereby is to make clear in this section the connection clear between the 
Hopf algebraic approach and the group theory.
The basic starting point is Weyl's character formula
\begin{align}
\ch(\Lambda) &= \sum_{w\in W} \epsilon(w)e^{w(\Lambda+\rho)}/
\sum_{w\in W} \epsilon(w)e^{w(\rho)}
\end{align}
where $\Lambda$ is the highest weight vector, $\rho$ is half the sum of
the positive roots and $W$ is the appropriate Weyl group with 
$\epsilon$ the sign of $w$. The Cartan 
classification of the simple complex classical Lie groups is given by the 
series $A_n$, $B_n$, $C_n$ and $D_n$, not to be confused with Schur 
function series. They correspond to the complexified versions of the 
groups $SU(n+1)$, $SO(2n+1)$, $Sp(2n)$ and $SO(2n)$. These
groups can be considered as subgroups of unitary groups $U(N)$ for
$N=n+1,2n+1,2n$ and $2n$. Denoting eigenvalues as $x_k=exp(i\phi_k)$ and 
$\bar{x}_k=x_k^{-1}$ one can write the eigenvalues of group elements in the
following way
\begin{align}
SU(n+1)  &&& x_1,x_2,\ldots,x_n\quad\text{with}\quad x_1x_2\ldots x_n=1& \nn
SO(2n+1) &&& x_1,x_2,\ldots,x_n,\bar{x}_1,\bar{x}_2\ldots\bar{x}_n,1&\nn
Sp(2n)   &&& x_1,x_2,\ldots,x_n,\bar{x}_1,\bar{x}_2\ldots\bar{x}_n&\nn
SO(2n)   &&& x_1,x_2,\ldots,x_n,\bar{x}_1,\bar{x}_2\ldots\bar{x}_n&
\end{align}
The connection to the group characters is obtained by inserting the
eigenvalues into the Weyl characterformula and interpreting the exponetials as
\begin{align}
e^{\lambda} &= x_1^{\lambda_1}x_2^{\lambda_2}\ldots x_n^{\lambda_n}
\qquad\lambda=(\lambda_1,\ldots,\lambda_n).
\end{align}
In the case of $U(n)$, the Weyl group is just the symmetric group (on $n$
letters). Hence the characters are labeled by partitions and the Weyl
character formula turns into the defining relation of the Schur functions. 
Let $\mu$ be the conjugacy class of the permutation. One finds
\begin{align}
\ch_\mu(\lambda) &= \sum_{w\in S_n} \epsilon(w)e^{w(\lambda+\rho).\mu}/
\sum_{w\in S_n} \epsilon(w)e^{w(\rho.\mu)},
\end{align}
where $\rho=(n-1,n-2,\ldots,1,0)$. Both numerator and denominator reduce to
determinants after inserting the $x_i$ (the denominator being the van der 
Monde determinant), and the quotient of the two alternating
functions is a standard construction of the Schur function \cite{macdonald:1979a}.
 
Let us introduce the standard notation for group characters
\begin{align}
U(n) &\qquad ch_\mu(\lambda) \,=\, \{\lambda\}(x)_n, \nn
O(n) &\qquad ch_\mu(\lambda) \,=\, [\lambda](x)_n, \nn
Sp(n) &\qquad ch_\mu(\lambda) \,=\, \langle\lambda\rangle(x)_n.
\end{align}
It is well known, that one has the following relations using $S$-function
series 
\begin{align}
U(n) \downarrow O(n) && \{\lambda\}(x) \,&=\, [\lambda/D](x)\nn
O(n) \uparrow U(n)   &&  [\lambda](x) \,&=\, \{\lambda/C\}(x)\nn
U(n) \downarrow Sp(n)&& \{\lambda\}(x) \,&=\, \langle\lambda/B\rangle(x)\nn
Sp(n) \uparrow U(n)  &&  \langle\lambda\rangle(x) \,&=\, \{\lambda/A\}(x)
\end{align}
In fact, this justifies the name \emph{branching operator} for the operators
which we had defined in (\ref{eq:bo}), at least in the cases $A,B$, and $C,D$.
Looking at the table (\ref{eq:series}), that $A,B$ are derived from the basic $L$ series by a
plethysm with $\{1^2\}$, reflecting the fact that symplectic groups
have antisymmetric bilinear forms as metric, while $C,D$ are related to the plethysm of the $L$ series with $\{2\}$,
reflecting this time the fact that orthogonal groups are based on symmetric
bilinear forms.

We noticed above, that formula (\ref{eq:3-70}) is valid for generating
functions of the form $\prod (1\pm f(x_i))$, for polynomial $f$. This was the
origin of these series being group like, and could be tied to a 1-cocycle
ensuring that the `branching operator' gave an algebra
homomorphism. In this sense only a trivial
branching process is involved (although the transformation involved
could be very complicated in detail). As an example, one might look at the series $V,W$
in table (\ref{eq:series}), which being plethysms by 
$(\{2\}-\{1^2\})\equiv p_{2}$
are group like, since $V=L(x_i^2)$, $W=L(x_i^2)^{-1}$. This may be
summarized in the statement
\myb{Lemma} All group like series $\Phi = \prod_{i}(1-f(x_{i}))^{s}$ (based on 
1-cocycles) induce trivial 
branchings, \textit{ie} branchings equivalent to $U(n)$.
\mye  \noindent
We noticed above, that 1-cocycles cannot induce nontrivial
2-cocycles, since by definition $\partial c_1 = e$.  For a product
deformation we would need a nontrivial 2-cocycle, at least a
2-coboundary, hence this outcome is in full accord with our
cohomological classification.  

Finally, the non-trivial series not based on
1-cocycles are no longer algebra homomorphisms, and one cannot
expect that the branching law for characters remains valid. 
Thus the $A,B$ and $C,D$ series are not group like.
However, the defect from being a
homomorphism, is fully compensated by the cliffordization \textit{wrt} the
Schur scalar product or its inverse, as shown in (\ref{eq:NL}), 
(\ref{eq:NL2}),  and (\ref{eq:4-84}) respectively.
This is a well known fact and describes the product rules of the
groups $Sp(n)$ and $O(n)$, which can equally be seen as the branching 
rules $Sp(n) \! \times \!Sp(n) \!\downarrow\!   Sp(n)$, $O(n) \! \times 
\! O(n) \!\downarrow\! O(n)$:
\myb{Theorem} (Newell-Littlewood \cite{newell:1951a,littlewood:1958a}) 
\begin{align}
Sp(n) &\qquad \langle\lambda\rangle {\scriptstyle{\otimes}} \langle \mu \rangle \,=\, 
\sum_{\zeta} \langle(\lambda/\zeta) \cdot (\mu/\zeta)\rangle \nn
O(n) &\qquad [\lambda] {\scriptstyle{\otimes}} [\mu] \,=\, 
\sum_{\zeta} [(\lambda/\zeta) \cdot (\mu/\zeta)] 
\end{align}
where $\cdot$ is the outer product of $S$-functions, $/$ the 
$S$-function quotient and ${\scriptstyle{\otimes}}$ is the Kronecker tensor 
product\footnote{The notation 
$\circ_{(\mid)*B}$ and $\circ_{(\mid)*D}$ for the cliffordized products 
is perhaps more suggestive.} of 
the universal characters. 
\mye \noindent
In fact, we found in Lemma 4.4. that the inverse of the branchings given by
the Newell-Littlewood theorem is given by the cliffordized products
$\circ_{(~.~\mid{\bf \antip}(~.~))*\Phi^{-1}}$, where the inverse series is 
employed, but also the convolutive inverse Schur scalar product 
$(~.~\mid {\sf \antip}(~.~))$. In this way,
that the cliffordization \textit{wrt} the Schur scalar product provides the
compensation for the fact that the series acting in the branching are not 
homomorphisms. This reads explicitly
\begin{align}
    \label{eq:circphiprod}
f\circ_{(~.~\mid~.~)*\Phi} g &= \Phi^{-1} (\Phi(f) \circ_{(~.~\mid~.~)}
\Phi(g)) \nn
&= \Phi^{-1} \sum (\Phi(f_{(1)}) \mid \Phi(g_{(1)})) 
\Phi(f_{(2)})\cdot \Phi(g_{(2)}).
\end{align}
This beautiful result immediately raises the following questions,
among others.  Is it possible to define other generic
cliffordizations, \textit{ie} inner products which are 2-cocycles, and
what are the corresponding series?  Also, while it is clear that
series such as $L(x_i^3)$ are still group like, it seems to be
questionable if series with three or more independent variables can
lead to an associative multiplication, \textit{ie} are based on a
2-cocycle.  Hence the question, what kind of `branching rule' can be
derived for series of the form say $\{3\}\circ L$, $\{2 1\}\circ L $,
or $\{1^3\}\circ L $?  In fact one awaits no `group' here, since all
classical groups are well known and exhausted by the above cases.

\section{Conclusions and analogy with quantum field theory}
In this paper we have given a synthesis of aspects of symmetric 
function theory from the viewpoint of underlying Hopf- and bi-algebraic 
structures. The focus has been on the explicit presentation of basic 
definitions and properties satisfied by the fundamental 
ingredients -- outer and inner products and coproducts, units and 
counits, outer antipode, Schur scalar product, skew product -- with 
ramifications for Laplace pairings, Kostka matrices, Sweedler cohomology and 
especially Rota cliffordizations. Our main result is that there is a 
rich class of associative deformations isomorphic to the standard outer product 
(but non-isomorphic as augmented algebras), and that the 
Newell-Littlewood product for symmetric functions of orthogonal and 
symplectic type is an associative deformation non-isomorphic to the 
outer product, derived from a 2-cocycle, up to a coboundary in the 
Sweedler sense. The conclusion of our analysis is the recognition 
that, even at this level, many familiar constructs from the rich 
theory of symmetric functions can be encapsulated by the organising power 
of the co-world of Hopf- and bi-algebras.

No attempt has been made to extend the analysis beyond 
the standard symmetric polynomials -- Hall-Littlewood, Macdonald, 
Jack, Kerov, MacMahon/vector, factorial/hypergeometric, \ldots symmetric 
functions should enter the Hopf framework at points where the 
structure admits natural generalisations.
For example, while we have considered the Schur scalar
product for the ring $\otimes_{\openZ}\Lambda$,  
the appearance of Hall-Littlewood and Macdonald symmetric functions shows
clearly the necessity for ring extensions and for the formerly symmetric
monoidal category to turn into a braided one. Thus $q$-Laplace 
expansions (in the context of a $q$-braided crossing) should play 
a crucial role in unveiling the nature of $q$-Kostka matrices 
and $q$-Littlewood-Richardson coefficients. Further extensions of 
$\openQ$ by irrationals obtain from evaluating the $q$ in the $q$-polynomials.
Such a ring $\openQ\otimes_{\openZ[v_1,v_2,\ldots]}\Lambda$ has a much more 
interesting cohomology group and therefore should bear a rich class of 
possible different cliffordizations (with non-cohomologous 2-cocycles). 

Finally a disclaimer should be made that in the present work no
attempt has been made to address deeper issues of symmetric function
theory such as outer and inner plethysms, the role of vertex
operators, and the fermion-boson correspondence.  For Hopf algebra
approaches to plethysm see \cite{scharf:thibon:1994a}; the
Littlewood-Richardson rule and the fermion-boson correspondence is
discussed in \cite{baker:1995a}.  For matrix elements of vertex
operators using composite supersymmetric $S$-functions see
\cite{jarvis:yung:1993}, and  \cite{baker:1996} for vertex operators for symmetric
functions of orthogonal and symplectic type. 
Relations to inner plethysm are given in
\cite{scharf:thibon:wybourne:1993}.

Several parallels between symmetric functions and combinatorial 
approaches to quantum field theory (QFT) have been alluded to in the 
text.  We conclude with an amplification of these points, which may 
provide further motivation for the programme outlined here.  An 
underlying cornerstone in combinatorics and in its application to 
QFT is the following: given a set of objects, called letters (or 
`balls' if we use a combinatorial notion), one is interested in the 
first instance in the relation of these objects (putting (weighted) 
balls together into boxes).  This level is given by the symmetric 
functions, or $\Tens[L]$ where $L$ is the letter `alphabet', 
hence the variables $\{ x_i\}$ (the balls).  The \textit{grading} of 
the tensor algebra imposes collections inhabited by symmetric 
functions of $n$-variables (putting $n$-balls into boxes).  Having 
objects and morphisms, we are ready to form a category.  In a 
second step, we are dealing with deformations of operations, which 
live in $\End \Tens[L]$.  This is asking for operations on operations, 
or more physically speaking `parameterization' of operations (putting 
boxes into packages).  Mathematically speaking we are dealing with a 
2-category.  In the theory of 
symmetric functions, Schur functions are used as symmetric 
functions, hence $s_\lambda \in \Tens[L]$ (boxes with balls) \emph{and 
at the same time} as polynomial functors (see \cite{macdonald:1979a} 
chapter I, appendix), \textit{ie}  as operators \emph{on} symmetric functions 
(packages collecting boxes).  It was to our knowledge Gian-Carlo Rota who 
made this explicit. Hence we are dealing with symmetric functions from 
$\Lambda(X) \cong \Tens[L]$ \emph{and} with endomorphisms living in 
$\Lambda^\otimes \cong \Lambda(X,Y,\ldots) \cong \Tens \Tens[L]$.  The 
process of moving up one step in a hierarchy or stack of categories, 
\textit{ie}  
moving from 1-categories to 2-categories (to $n$-categories) is  
\emph{categorification} \cite{baez:dolan:1998a}. 

The triply iterated structure alluded to here is perfectly mirrored in
the process of second quantisation (coordinates $\rightarrow$
wavefunctions $\rightarrow$ functionals), and supports the suggestion
that some of the machinery of quantum field theory can be captured at
the combinatorial level.  Recall for instance the role of Kostka
matrices (Lemma 2.10 above) in counting
column and row sum matrices.  As mentioned, in quantum field theory
the \emph{fields}\footnote{Or currents, in a generating functional
approach; formally, the label $\phi(x)$ should be associated with an
element ${[}\phi|x{]}$ of a letter-place alphabet in the sense of Rota
\cite{grosshans:rota:stein:1987a}.} have to be considered as the
variables `$x$', and hence the double Kostka matrices $M(h,m)$ or
$M(e,m)=[a_{ij}]$ appear as exponents in expressions like
\begin{align}
\sum\sum\prod\prod \left(\phi_i(x)\vert\phi_j(y)\right)^{a_{ij}},
\end{align}
where the $a_{ij}$ are from $M(e,m)$ for fermions and from $M(h,m)$ for 
bosons, and the `scalar products' have to be replaced by suitable propagators.
Details may be found in \cite{brouder:schmitt:2002a} or any book on quantum 
field theory dwelling on renormalization.

It was argued \textit{eg} in 
\cite{fauser:1996c,fauser:stumpf:1997a,fauser:1998a,%
fauser:2001e,fauser:2002c}, that the structure of a quantum field
theory is governed by two structures: the quantization, a bilinear
form of the opposite symmetry type from that of the fields, and the propagator, a
bilinear form of the same symmetry type.  A detailed account of these
findings in quantum field theory in Hopf algebraic terms is given in
\cite{brouder:fauser:frabetti:oeckl:2002a}, where also the cohomology
is used as \emph{classifying principle}.  The same holds true 
for symmetric function theory, and indeed a broader analogy between the algebraic structures present in the 
symmetric functions and QFT comes by considering the deforming products.
As we defined the branching 
operators, they may be seen as \textit{`active'} endomorphisms transforming any element 
of $\Lambda$ into another such element, especially a Schur function into 
a series $\{\lambda\}/\Phi$, \textit{eg}
$s_{(21)}/M  = \sum_{m\in M} s_{(21)/m} =
s_{(21)} + s_{(2)} + s_{(1^2)} + s_{(1)}$. 
Due to our construction, these transformations are invertible. Writing this in
a more Hopf algebraic flavour, we had:
$ \Phi(s_{\lambda}) = \sum_{\lambda} 
\phi(s_{\lambda(1)})s_{\lambda(2)}$, and
$\Phi^{-1}(s_{\lambda}) 
= \sum_{\lambda} \phi^{-1}(s_{\lambda(1)})s_{\lambda(2)}$.
Looking at the structure of QFT, it was noted
\cite{fauser:2001b,brouder:2002a,fauser:2002c,%
brouder:fauser:frabetti:oeckl:2002a} that this formula is nothing but
a Wick transformation from normal- to time-ordered field operator
products.  Considering the $s_{\lambda}$ as the `normal ordered'
basis, the above formula computes the `time ordered' expression
$s_\lambda/M$ in the `normal ordered' basis.  Of course, time and
normal ordering is just a name tag in the theory of symmetric
functions.  However, this opens up a second, \textit{`passive'}, perspective,
which we believe to be a new result in the theory of symmetric
functions.  Define a new \textit{dotted} `outer product', denoted by
$:$ for the moment\footnote{Dotted wedge products were introduced
\textit{eg} in \cite{fauser:ablamowicz:2000c,fauser:2002c}; in the
present context the `dotted' product is of course that developed in \S
4; see for example (\ref{eq:circphiprod}) above.}.  The non augmented
outer product algebras $(\Lambda,\cdot)$ and $(\Lambda,:)$ are
isomorphic, hence there exists an isomorphism $s_{\lambda}\cdot \ldots
\cdot s_{\mu} \mapsto s_{\lambda}: \ldots : s_{\mu}$.  With the
artificial terminology, borrowed from quantum field theory, this
amounts to saying that the algebra isomorphism transforms the `normal
ordered' outer product $\cdot$ into the `time ordered' dotted outer
product $:$.  In quantum field theory however an additional structure
must be taken into account which destroys the isomorphy -- namely the
unique vacuum described by the counit.  The same happens to be true
for symmetric functions, where the counit is the evaluation of the
symmetric functions at $x_i=0$ for all $x_i$.  This evaluation is
different in time and normal ordered expressions as we demonstrated in
(\ref{eq:3-69}) above.

As claimed in the introduction, symmetric functions may serve as a 
laboratory for quantum field theory.  Given the importance of symmetry 
computations in multi-particle quantum systems, as in quark models, 
the nuclear shell-model, the interacting boson and the vibron model, 
spectrum generating groups, as well as exactly solvable models in 
quantum field theory and statistical mechanics, and especially two 
dimensional systems and the fermion - boson correspondence, it is 
perhaps not surprising that that a very close analogy can be found.
As far as symmetric function theory itself is concerned, questions 
raised by the present study include for example a deeper Sweedler-
cohomological classification of ring extensions, and an associated 
classification of branching and product rules for the therewith-attached 
(quantum) (affine) (non-compact) Lie (super) groups? Even the un-bracketed words 
in this list of attributes come under scrutiny in the light of 
the above discussion of the possible role of Littlewood's series $\{3\}\circ  L$, 
$\{2 1\}\circ L $, or $\{1^3\}\circ L $.
Applications to extended, possibly 
non-associative, algebraic structures which may relate to 
compositeness may be implied by the present framework. 

\paragraph{Acknowledgement:} The authors would like to thank Jean-Yves Thibon 
for sending reprints of his work. BF gratefully acknowledges the support 
of an ARC Research Fellowship during his visit at the University of Tasmania
at Hobart, May--July 2003 (project DP0208808).

\begin{appendix}
\section{Hopf algebras versus $\lambda$-rings}
\def\theequation{\Alph{section}-\arabic{equation}}

\subsection{Definition of $\lambda$-ring}

Since we used in this paper partly $\lambda$-ring notion, it might well serve
the reader to have the definitions around and to explore the relations
between $\lambda$-rings and Hopf algebras a little further.
Let ${\bf R}$ be a commutative unital ring. To form a $\lambda$-ring the
following additional requirements are imposed.
\myb{Definition}
A $\lambda$-ring is a ring ${\bf R}$ supplemented with the following structure
maps ($r,s\in {\bf R}$): 
\begin{align}
\lambda^0(r) &= 1 \nn
\lambda^1(r) &=r \nn
\lambda^n(1) &= 0\quad \forall n>1\nn
\lambda^n(r+s) &= \sum_{p+q=n} \lambda^{p}(r)\lambda^{q}(s) \nn
\lambda^n(rs)  &= {\sf P}_n(\lambda^1(r),\ldots,\lambda^n(r);
                   \lambda^1(s),\ldots,\lambda^n(s) ) \nn
(\lambda^n\circ\lambda^m)(r)\,=\,\lambda^n(\lambda^m(r)) 
             &= {\sf P}_{n\cdot m}(\lambda^1(r),\ldots,\lambda^{n\cdot m}(r)),
\end{align}
where ${\sf P}_n$ and ${\sf P}_{n\cdot m}$ are certain universal polynomials
with integer coefficients \cite{atiyah:tall:1969a,knutson:1973a}.
\mye \noindent 
In fact we can give the polynomials ${\sf P}_n$ and ${\sf P}_{n\cdot m}$ in
the case of symmetric functions as follows. Let $x=x_1+x_2+x_3+\ldots$ be an
\emph{alphabet} of grade 1. Hence we will consider the $\lambda$-ring
$\Lambda(x)$ generated by this alphabet $x$. Introduce a second alphabet 
$y=y_1+y_2+y_3+\ldots$ and the elementary symmetric functions for these two
sets of indeterminants
\begin{align}
(1+e_1(x)t+e_2(x)t^2+\ldots ) &= \prod_{i} (1+x_i\,t) \nn
(1+e_1(y)t+e_2(y)t^2+\ldots ) &= \prod_{i} (1+y_i\,t)
\end{align}
Then ${\sf P}_{n}(e_1(x),e_2(x),\ldots e_n(x),e_1(y),e_2(y),\ldots,e_n(y))$
is defined to be the coefficient of $t^n$ in
$\prod_{i,j}(1+x_iy_jt)$. Similarly ${\sf P}_{n\cdot m}(e_1(x),e_2(x),
\ldots,e_{n\cdot m}(x))$ is the coefficient of $t^{n}$ in the following 
product
$\prod_{1\le i_1\le i_2 \ldots \le i_d \le q}(1+x_{i_1}x_{i_2}\ldots
x_{i_d} t)$. Neither polynomial depends on $q,r$ if the number of
variables is sufficiently large, and are called universal polynomials.

The well known relation to formal power series is as follows
\cite{hazewinkel:1978a}. Let ${\bf R}$ be any commutative unital ring, 
then there is a functor $\Lambda : {\bf ring}\rightarrow {\bf ring}$ 
which assigns to ${\bf R}$ the universal $\lambda$-ring $\Lambda({\bf R})$. 
Consider formal power series of the form
\begin{align}
f(t) &= 1 +\sum_{i\ge1} r_i\,t^i \quad g(t)\,=\, 1+\sum_{i\ge1}s_i\,t^i 
\end{align}
and define addition and multiplication in $\Lambda({\bf R})$ as 
\begin{align}
f(t)+_{\!\Lambda}\,g(t) &= 1+ \sum_{i\ge1}\left(
\sum_{n+m=i}r_n\,s_m\right)\,t^i\nn
f(t)\cdot_{\!\Lambda}\,g(t) &= 1+ \sum_{i\ge1}
{\sf P}_i(r_1,\ldots,r_i;s_1,\ldots,s_i)\,t^i
\end{align}
where ${\sf P}_i$ is the polynomial appearing in the definition of the
$\lambda$-ring. Furthermore, for any ring homomorphism $h : {\bf R}
\rightarrow {\bf S}$ one defines $H : \Lambda({\bf R}) \rightarrow 
\Lambda({\bf S})$ as
\begin{align}
H(1+\sum_{i\ge1} r_it") &= 1+\sum_{i\ge1} h(r_i)t^i 
\,=\, 1+\sum_{i\ge1} s_i\,t^i 
\end{align}
where the $r_i\in {\bf R}$ and the $s_i\in {\bf S}$. This turns 
$\Lambda$ into an endofunctor on {\bf ring}. 
The action of $\lambda^i$ on elements of $\Lambda({\bf R})$ is defined to be
\begin{align}
\lambda^i(f(t)) &= 1+\sum_{j\ge1} {\sf P}_{i \cdot j}(r_1,\ldots,r_{ij})\,t^j
\end{align}
where the ${\sf P}_{ij}$ are the polynomials of the definition of the
$\lambda$-ring.

Further important $\lambda$-ring operations are the Adams operations, defined
as 
\begin{align}
\psi^1(r) &= r \nn
\psi^n(\psi^m(r)) &= \psi^m(\psi^n(r)) \,=\, \psi^{n\cdot m}(r)
\end{align}
Of course, from the definition of the $\lambda$-ring and this relations one
reads of that Adams operations are connected with composition or plethysm
and the power sum basis.

The $\lambda$-maps can be used to form a ring-map from ${\bf R}$ to 
$\Lambda({\bf R})$ which assigns to every $r$ a formal power series
\begin{align}
\lambda_t &: {\bf R} \rightarrow \Lambda({\bf R}) \nn
\lambda_t(r) &= \sum_{i\ge0} \lambda^i(r)\,t^i \,=\, 1 + \sum_{i\ge1}
\lambda^i(r)\,t^i \,=\, 1+\sum_{i\ge1} r_i\, t^i 
\end{align}

With this in mind, it is easily seen how the translation table in Macdonald
\cite{macdonald:1979a}, p. 18 occurs. One obtains the translations
\begin{align}
x     &= x_1+x_2+\ldots +x_n+\ldots \nn
e_r   &\leftrightarrow \lambda^r(x)\qquad\text{$r$th exterior power} \nn 
E(t)  &\leftrightarrow \lambda_t(x) \nn
h_r   &\leftrightarrow \sigma^r(x)\qquad\text{$r$th symmetric power} \nn  
H(t)  &\leftrightarrow \sigma_t(x) \,=\, \lambda_{-t}(-x) \nn
p_r   &\leftrightarrow \psi^r(x) \qquad\text{Adams operations} \nn
P(t)  &\leftrightarrow \lambda^{-1}_{-t}(x)\frac{{\sf d}}{{\sf dt}}
        \lambda_{-t}(x) \,=\, \frac{{\sf d}}{{\sf dt}} \log \lambda_{-t}(x)
\end{align}

Rota and collaborators used letter-place super algebras for their works in
invariant theory and combinatorics \cite{grosshans:rota:stein:1987a}, which is
related to the $\lambda$-ring formalism as follows. Let $L$ be an alphabet of
possibly signed letters --we assume positive letters to avoid sign problems.
As can be deduced from \cite{rota:stein:1994a,rota:stein:1994b}, the theory of
symmetric functions is derived from a letter place algebra of a single letter
$x$, which therefore has to be considered in terms of $\lambda$-ring
structures. In fact, the ring $\Lambda$ of symmetric functions is the free
ring underlying the $\lambda$-ring $\Lambda(X)$ in a singe variable,
\cite{macdonald:1979a}, p. 17. In this sense, Rota's plethystic algebra 
$\Pleth[L]$ is concerned with those parts of the $\lambda$-ring structure which 
are related with ${\sf P}_{nm}$.  

\subsection{Hopf algebraic aspects of $\lambda$-rings}

Note, that one can assign to a generating function like $H(t)$ a 
Toeplitz matrix, that is a band matrix with entries $h_{ij}=[h_{i-j}]$. This
morphism from generating functions to matrices turns the pointwise product
of ordinary polynomial sequence generating functions (opgf, \cite{wilf:1990a})
into the matrix product of Toeplitz matrices.
\begin{align}
f(t)\cdot g(t) &\cong 
\left[\begin{array}{ccccc}
f_0 & f_1 & f_2 & f_3 & \ldots \\
0   & f_0 & f_1 & f_2 & \ldots \\
0   &   0 & f_0 & f_1 & \ldots \\
0   &   0 &   0 & f_0 & \ldots \\
\vdots & \vdots & \vdots & \vdots & \ddots 
\end{array}\right] \circ
 \left[\begin{array}{ccccc}
g_0 & g_1 & g_2 & g_3 & \ldots \\
0   & g_0 & g_1 & g_2 & \ldots \\
0   &   0 & g_0 & g_1 & \ldots \\
0   &   0 &   0 & g_0 & \ldots \\
\vdots & \vdots & \vdots & \vdots & \ddots 
\end{array}\right]
\,=\,
 \left[\begin{array}{ccccc}
h_0 & h_1 & h_2 & h_3 & \ldots \\
0   & h_0 & h_1 & h_2 & \ldots \\
0   &   0 & h_0 & h_1 & \ldots \\
0   &   0 &   0 & h_0 & \ldots \\
\vdots & \vdots & \vdots & \vdots & \ddots 
\end{array}\right]
\end{align}
where $h_i = \sum_{n+m=i} f_n\,g_m$ and $h_i=0$ if $i<0$. Now, this can be 
easily recast in terms of a convolution product of a coalgebra and an 
algebra map. Let 
\begin{align}
\delta(t) &= t\otimes t,\qquad \mu(t^r\otimes t^s) \,=\, t^{r+s}
\end{align}
then one finds using maps $F: t \rightarrow \sum f_i\, t^i$ and $G$ 
analogously,
\begin{align}
(F*G)(t) &= \mu\circ(F\otimes G)\circ \delta(t) \,=\, F(t)\cdot G(t) \nn
         &= \sum_{n+m=i} f_ng_m t^i \,=\, \sum_i\sum_r f_rg_{i-r}t^i.
\end{align}
Hence the pointwise product of generating functions can be understood as a
convolution algebra made from a coalgebra-algebra pair. Note that the
convolution \emph{product} is related to $\lambda$-ring \emph{addition}. 
Moreover, one can dualize this approach defining a suitable coproduct 
acting on the coefficients of the generating functions. Therefore we 
define
\begin{align}
\Delta(e_n) &= \sum_{r=0}^n e_r\otimes e_{n-r} \nn
\Delta(e_n)&= \theta^{-1}(\Delta(e_n)(X+Y))
\,=\, \sum_{r=0}^n \theta^{-1}(e_r(X)e_{n-r}(Y)) 
\end{align}
where we have used the map $\theta$ from section (\ref{sec:2.5}) to make 
the connection between tensor formulations and formulation in 
$\lambda$-rings.

\noindent
{\bf Aside:} The crucial property of the maps $\lambda^i$ is that they are
similar to sequences of binomial polynomials. These binomial sequences have 
been studied by Rota and collaborators for quite a while \cite{rota:etal:1975a}. 
It is shown there, that every shift invariant operator gives rise to a set
of polynomials in such a way that they are Appell 
or Scheffer sequences (\textit{loc cit} p.58), hence fulfilling the property
\begin{align}
p_n(x+y) &= \sum_{s+r=n} \left( \! \begin{array}{c} n \\ r \end{array} 
\! \right)  p_s(x)p_r(y) 
\end{align} 
This relation, including the binomial coefficients, is related to exponential
generating functions (egf), and hence to the Adams operations and not directly
to the $\lambda^i$ maps. But it can be shown, that there are shift operators
$E^a$ which act as
\begin{align}
E^a\,p_n(x) &= p_n(x+a)
\end{align}
which is exactly the case for the $\lambda_{1}$ and $\sigma_1$ series in 
$\lambda$-ring notation 
\begin{align}
F(X) /\lambda_{1} &= F(X-1),\qquad F(X)/\sigma_{1} \,=\, F(X+1).
\end{align}
It would be extremely interesting to have explicitly the details of this
relation, which involves umbral calculus and umbral composition. 
The vertex operator $\Gamma(1)\circ F(X) = F(X)/{\sigma_{1}}/{\lambda_{1}}$
embodies in combinatorial terms the principle of inclusion and exclusion
(PIE), which is understood in Hopf algebraic terms via the theory of
\emph{species} developed by Joyal.

%
%
\end{appendix}

{

}
\noindent
{\sc 
Bertfried Fauser, Universit\"at Konstanz, Fachbereich Physik, Fach M678,
D-78457 Konstanz, Germany, {\small\tt Bertfried.Fauser@uni-konstanz.de}\\
Peter D. Jarvis, University of Tasmania, School of Mathematics and Physics,
GPO Box 252C, 7001 Hobart, TAS, Australia,  
{\small\tt Peter.Jarvis@utas.edu.au} \\
Report Number UTAS-PHYS-2003-02}
\newpage
\tableofcontents
\end{document}